\documentstyle[prc,aps,floats,epsfig,12pt]{revtex}
\begin{document}

\newcommand{\be}{\begin{equation}}
\newcommand{\ee}{\end{equation}}

%\documentstyle[aps,epsf,amsbsy,12pt]{revtex}
%\documentstyle[epsf,amsbsy,12pt]{article}
%\begin{document}

%\renewcommand{\baselinestretch}{1.66}

\title{Chiral Phase Transition within Effective Models
with Constituent Quarks}

\author{O. Scavenius$^1$, \'A. M\'ocsy$^2$, I.N. Mishustin$^{1,3,4}$ and 
D.H. Rischke$^5$}
\address{$^1$The Niels Bohr Institute, Blegdamsvej 17\\
DK--2100 Copenhagen \O, Denmark\\
$^2$School of Physics and Astronomy, University of Minnesota\\
Minneapolis, MN 55455, USA \\
$^3$The Kurchatov Institute, Russian Research Center\\
Moscow 123182, Russia \\
$^4$Institut f\"ur Theoretische Physik, J. W. Goethe Unerversit\"at, \\
D-60054 Frankfurt am Main, Germany \\
$^5$RIKEN-BNL Research Center, Brookhaven National Laboratory, \\
11973 Upton, New York, USA\\
}

\maketitle

\begin{abstract}

We investigate the chiral phase transition at nonzero temperature $T$ and
 baryon-chemical potential $\mu_B$ within the framework of the linear
sigma model and the Nambu--Jona-Lasinio model. For small bare quark masses 
we find in both models a smooth crossover
transition for nonzero $T$ and $\mu_B=0$ and a first order 
transition for $T=0$ and nonzero $\mu_B$. We calculate explicitly 
the first order phase transition line and spinodal lines in the $(T,\mu_B)$ plane. As expected
they all end in a critical point. We find that, in the linear sigma model, the sigma mass goes to zero at the critical point. This is in
contrast to the NJL model, where the sigma mass, as defined in the random phase approximation, does not vanish. We also compute the
adiabatic lines in the $(T,\mu_B)$ plane. Within the models studied here, the critical point does not serve as a ``focusing'' point in the adiabatic
expansion.
\end{abstract}

\section{Introduction} 
Chiral symmetry is spontaneously broken in the QCD vacuum.
Lattice QCD simulations at nonzero
temperature $T$ and zero baryon-chemical potential $\mu_B$  
indicate that chiral symmetry is restored above a temperature
$T \sim 150$ MeV \cite{lattice}.
Such temperatures are believed to be created in nuclear collisions at 
ultra-relativistic energies. Consequently, a phase where
chiral symmetry is transiently restored may be formed in these collisions.
%\cite{harrismuller}. 
 The subsequent expansion cools the system and takes it to the final
hadronic state, where chiral symmetry is again spontaneously broken.

It is important to determine the order of the chiral transition, as
this influences the dynamical evolution of the system.
For instance, it has been shown that a first order transition may lead
to a deflagration wave and to a ``stall'' in the expansion of 
the system \cite{mgdhr}.
It has also been shown that a
first order transition in rapidly expanding matter may manifest itself by 
strong non-statistical fluctuations due to droplet formation \cite{igor}.
In the case of strong supercooling it may lead
to large fluctuations due to spinodal decomposition \cite{ms,oa}.
In a second order phase transition one may expect the appearance of
critical fluctuations due to a large correlation length \cite{SRS}. 
Experimentally, large-acceptance detectors are now able to measure average as well as event-by-event observables,
which in principle allow to distinguish between scenarios with a 
first order, a second order, or merely a crossover type of
a phase transition.
%\cite{ste}.

Theoretically, the QCD phase diagram in the $(T,\mu_B)$ plane has 
recently received much attention (see \cite{SRS,Halasz,Klevansky,ARW}).
QCD with $N_f=2$ flavors of massless quarks has a global 
$SU(2)_L \times SU(2)_R$ symmetry. 
This symmetry is spontaneously broken in the QCD vacuum, such that
the order parameter $\phi^{ij}
\sim \langle \bar{q}_L^i q_R^j \rangle$ acquires a
non-vanishing expectation value, where $q^i$ is the quark
field ($i,j$ are the flavor indices). At zero baryon-chemical potential, 
the effective theory for this order parameter is the same as the
$O(4)$ model which has a second order phase transition. Therefore, by 
universality
arguments \cite{rob}, the chiral transition in $N_f = 2$ QCD is likely to be of second order at $\mu_B = 0$.
Nonzero quark masses introduce a term in the QCD Lagrangian which
explicitly breaks chiral symmetry. Then, the second order transition becomes
crossover. 

At nonzero baryon-chemical potential, it is more difficult to infer the 
order of the chiral transition from universality arguments
\cite{Hsu}. One commonly resorts
to phenomenological models to describe the chiral transition in this case. 
Depending on the parameters of these models, they
predict a first order, a second order, or a crossover transition.
However, if there is a second order phase transition
for $\mu_B=0$ and nonzero $T$ and a 
first order transition for small $T$ and nonzero $\mu_B$, 
then there exists a tricritical point in the $(T,\mu_B)$ plane where
the line of first order phase transitions meets the
line of second order phase transitions. For nonzero quark masses,
this tricritical point becomes a critical point. 

It has recently been proposed \cite{SRS} that 
this point could lead to 
interesting signatures in heavy-ion collisions at 
intermediate energies,
if the evolution went through or close to this critical
point. At this point, susceptibilities (e.g.\, the heat capacity) 
diverge, and the order parameter
field becomes massless and consequently fluctuates strongly, which
could be detected in event-by-event observables.

In this paper we investigate the thermodynamics of two
popular models of chiral dynamics, 
the linear sigma model coupled
to quarks \cite{linsig}, and the Nambu--Jona-Lasinio (NJL) model \cite{NJL}.
Both models are tuned to reproduce correctly properties
of the physical vacuum.
Our goal is to study the chiral transition and to verify the existence of the critical point at nonzero
chemical potential and temperature. We also study the
behavior of isentropes in the vicinity of the phase transition line
in the ($T,\mu_B$) plane. These results can then be used
in dynamical simulations to confront the predictions of \cite{mgdhr,igor,ms,oa,SRS} with
experimental data.

The structure of the paper is as follows.
In Section II we study the thermodynamics of the 
linear sigma model coupled to quarks. This part of the 
paper is an extension of our previous study in ref. \cite{mocs}.
 In Section III we do the same 
for the NJL model.
Section IV presents numerical results. We conclude
in  Section V with a summary of our results.
Our units are $\hbar=c=k_B =1$, the metric tensor is
$g^{\mu \nu} = {\rm diag} (+,-,-,-)$.

\section{Thermodynamics of the Linear Sigma Model}

The Lagrangian of the linear sigma model with quark degrees of freedom 
reads
\begin{equation}
{\cal L}=\overline{q} \left[ i\gamma ^{\mu}\partial _{\mu}-
g(\sigma +i\gamma _{5}\vec{\tau} \cdot \vec{\pi} )\right] q
+ \frac{1}{2} \left(\partial _{\mu}\sigma \partial ^{\mu}\sigma + 
\partial _{\mu}\vec{\pi} \cdot \partial ^{\mu}\vec{\pi}\right)
-U(\sigma ,\vec{\pi}) \,\, ,
\label{sigma}
\end{equation}
where the potential is
\begin{equation}
U(\sigma ,\vec{\pi} )=\frac{\lambda^{2}}{4} \left(\sigma ^{2}+\vec{\pi} ^{2} 
-{\it v}^{2}\right)^{2}-H\sigma \,\, .
\end{equation}
Here $q$ is the light quark field $q=(u,d)$. The
scalar field $\sigma$ and the pion field 
$\vec{\pi} =(\pi _{1},\pi _{2},\pi _{3})$ together form a chiral field 
$\Phi =(\sigma,\vec{\pi})$. This Lagrangian is invariant under chiral 
$SU(2)_L \times SU(2)_R$ transformations if the explicit symmetry breaking 
term $H\sigma $ is zero. The parameters of the Lagrangian are usually 
chosen such that the chiral symmetry is spontaneously broken in the 
vacuum and the expectation values of the meson fields are 
$\langle\sigma\rangle ={\it f}_{\pi}$ and $\langle\vec{\pi}\rangle =0$, 
where ${\it f}_{\pi}=93$ MeV is the pion decay constant. 
The constant $H$ is fixed by the PCAC relation which gives 
$H=f_{\pi}m_{\pi}^{2}$, where $m_{\pi}=138$ MeV is the pion mass. 
Then one finds $v^{2}=f^{2}_{\pi}-m^{2}_{\pi}/\lambda^{2}$. 
The coupling constant $\lambda^2$ is determined by the sigma mass, 
$m_\sigma^2=2\lambda^2 f^2_\pi + m^2_\pi$, 
which we set to 600 MeV, yielding $\lambda^{2} \approx 20$. 
The coupling constant $g$ is usually fixed by the requirement that the 
constituent quark mass in vacuum, $M_{\rm vac}=gf_\pi$, is about 
$1/3$ of the nucleon mass, which gives $g \simeq 3.3$. 

Let us consider a system of quarks and antiquarks in thermodynamical
equilibrium at temperature $T$ and quark chemical
potential $\mu\equiv \mu_B/3$.
The grand partition function reads:
\begin{equation}
{\cal Z}={\rm Tr} \, \exp \left[-\left(\hat{\cal H}-\mu \hat{\cal N}
\right)/T \right]
= \int{\cal D}\bar{q}\, {\cal D}q\, {\cal D}\sigma\, {\cal D}\vec{\pi}\;
\exp\left[\int_x \left({\cal L}+\mu \, \bar{q} \gamma^0 q\right) \right]
\,\, .
\end{equation}
Here $\int_x \equiv i \int_0^{1/T} dt \int_V d^3 {\bf x}$, where $V$
is the volume of the system.
We adopt the mean-field approximation, replacing
$\sigma$ and $\vec{\pi}$ in the exponent by their expectation values.
Then, up to an overall normalization factor:
\begin{eqnarray}
{\cal Z} & = & \exp\left(-\frac{VU}{T}\right) \;
\int{\cal D}\bar{q}\, {\cal D}q\; 
\exp\left\{ \int_x \bar{q} \left[ i\gamma ^{\mu}\partial _{\mu}-
g(\sigma +i\gamma _{5}\vec{\tau} \cdot \vec{\pi})\right]q + 
\mu \bar{q} \gamma^0 q \right\} \nonumber \\
& = & \exp\left(-\frac{VU}{T}\right)
{\rm det}_p \left\{ \left[ p_{\mu}\gamma^{\mu} + \mu \gamma^0 -
g(\sigma +i\gamma_{5}\vec{\tau} \cdot \vec{\pi}) \right]/T\right\}\,\, .
\end{eqnarray}

All thermodynamical quantities can be obtained from the grand canonical
potential
\begin{equation}
\Omega (T,\mu)=-\frac{T\ln {\cal Z}}{V}=
U(\sigma,\vec{\pi})+ \Omega_{q\bar{q}},
\label{ptotsig}
\end{equation}
where the quark and antiquark contribution reads:
\begin{equation}
\Omega_{q\bar{q}}(T,\mu) = -\nu_{q} \int \frac{d^3{\bf p}}{(2\pi)^3}
\left\{E+T\, \ln \left[1+\exp \left(\frac{\mu-E}{T}\right) \right]
 +   T\, \ln \left[1+\exp \left(\frac{-\mu-E}{T}\right) \right] \right\}\,\, .
\label{pqq}
\end{equation}
Here, $\nu_q = 2 N_c N_f = 12$ is the number of internal
degrees of freedom of the quarks, $N_c = 3$, and 
$E=\sqrt{p^2+M^2}$ is the valence quark and antiquark energy.
 The constituent quark (antiquark) mass, $M$, is defined to be:
\begin{equation}
M^2=g^2(\sigma^2+\vec{\pi}^2)\,\, .
\label{qmass}
\end{equation}
The divergent first term in eq. (\ref{pqq}) comes from the negative energy
states of the Dirac sea. As follows from the standard renormalization
procedure it can be partly absorbed in the coupling constant 
$\lambda^2$ and the constant $v^2$. However, a logarithmic correction
from the renormalization scale remains and is neglected
in the following. Similar logarithmic terms are explicitly included
in calculations within the NJL model (see below). Therefore one can use the 
comparison of these two models to conclude about the importance of these
corrections.

After integrating eq. (\ref{pqq}) by parts the contribution
 of valence quarks and antiquarks can
 be rewritten as 
\begin{equation}
P_{q\bar{q}}(T,\mu) = \frac{\nu_q}{6\pi^2}\int^{\infty}_0 dp \frac{p^4}{E}\left [
n_q(T,\mu)+n_{\bar{q}}(T,\mu)\right ],
\end{equation}
where $n_q$ and $n_{\bar{q}}$ are the quark and antiquark occupation
numbers,
\begin{equation} \label{occupationnumbers}
n_q(T,\mu)=\frac{1}{1+\exp[(E-\mu)/T]}\,\, ,\hspace{1cm}
n_{\bar{q}}(T,\mu)=n_q(T,-\mu)\,\, .
\label{occunum}
\end{equation}

The baryon-chemical potential is determined by the net baryon
density
\begin{equation}
n_B = -\frac{1}{3}\frac{\partial \Omega}{\partial \mu} =
\frac{\nu_q}{6\pi^2}\int p^2dp [n_q(T,\mu)-n_{\bar{q}}(T,\mu)].
\end{equation}
The net quark density is obviously $n=3n_B$.
The values for the $\sigma$ and $\vec{\pi}$ fields and thereby the 
quark masses in eq.\ (\ref{qmass}) are obtained by
minimizing $\Omega$ with respect to $\sigma$ and $\vec{\pi}$,
\begin{equation}
\frac{\partial \Omega }{\partial \sigma}=\lambda^2(\sigma^2+\vec{\pi}^2-v^2)
\sigma-H+g\rho_{s}=0\,\, ,
\label{smmass}
\end{equation}

\begin{equation}
\frac{\partial \Omega }{\partial \pi_i}=\lambda^2(\sigma^2+\vec{\pi}^2-v^2)
\pi_i+g\rho_{ps,i}=0\,\, .
\label{pmmass}
\end{equation}
The scalar and pseudoscalar densities can be expressed as:
\begin{equation} \label{scalardensity}
\rho_s=\langle\bar{q}q\rangle=
g \sigma \nu_q \int \frac{d^3{\bf p}}{(2\pi)^3}
\frac{1}{E}[n_q(T,\mu)+n_{\bar{q}}(T,\mu)] \,\, ,
\label{scaldens}
\end{equation}
\begin{equation}
\vec{\rho}_{ps}=\langle\bar{q}i\gamma_5\vec{\tau}q\rangle
=g \vec{\pi} \nu_q \int \frac{d^3{\bf p}}{(2\pi)^3}
\frac{1}{E}[n_q(T,\mu)+n_{\bar{q}}(T,\mu)] \,\, .
\end{equation}
The minima of $\Omega$ defined by eqs. (\ref{smmass}), (\ref{pmmass})
correspond to the stable or metastable states of matter in thermodynamical 
equilibrium where the pressure is $P=-\Omega_{min}$.
The $\sigma$ and pion masses are determined by the curvature of 
$\Omega$ at the global minimum:
\begin{equation}
M^2_{\sigma} = \frac{\partial^2\Omega}{\partial \sigma^2}\;\;\; ,\;\;\;\;
M^2_{\pi_i} = \frac{\partial^2\Omega}{\partial \pi_i^2}\,\, .
\label{mmass}
\end{equation}
Explicitly they are given by the expressions
\begin{eqnarray}
M^2_{\sigma}& =& m^2_\pi+\lambda^2\left (3\frac{M^2}{g^2}-f^2_\pi \right )
 \nonumber \\
& + &  g^2\, \frac{\nu_q}{2\pi^2} \int dp\,p^2\;
\left[ \frac{p^2}{E^3}
\left( \frac{1}{1+\exp[(E+\mu)/T] }+\frac{1}{1+\exp[(E-\mu)/T]}\right)
\right.
\nonumber \\
& - &\left.  \frac{M^2}{T\,E^2} \left(\frac{1}{2(1+\cosh[(E+\mu)/T])}+
\frac{1}{2(1+\cosh[(E-\mu)/T])} \right) \right]\,\, ,
\end{eqnarray}
\begin{equation}
M^2_{\pi}=m^2_{\pi}+\lambda ^2(\frac{M^2}{g^2}-f^2_\pi)+g^2\, \frac{\nu_q}{2\pi^2}
\int dp\, p^2 \; \frac{1}{E}\left[ \frac{1}{1+\exp{[(E+\mu)/T]}}
+ \frac{1}{1+\exp{[(E-\mu)/T]}}\right].
\end{equation}
Here we have set the expectation value of the pion field to zero,
$\vec{\pi} = 0$, thus $M^2=g^2\sigma^2$.
This version of the sigma model was used earlier in ref. \cite{mocs} for 
 thermodynamical calculations at nonzero $T$ and $\mu=0$, and at nonzero $\mu$
and $T=0$. Some useful formulae for the case of small quark mass are given
in the Appendix.

\section{Thermodynamics of the NJL model}

The NJL model has been widely used earlier for describing 
hadron properties  and the chiral phase transition \cite{weise,klevanskyr}. 
The simplest version of the model including only
scalar and pseudoscalar 4-fermion interaction terms is given 
by the Lagrangian \footnote{As demonstrated in ref. \cite{mishusat}
, the inclusion of the vector-axialvector terms may
change significantly the parameters of the chiral phase transition,
in particular, the position of the critical point. But this does not change the
qualitative conclusions of the present paper.}:
\begin{equation}
{\cal L}=\bar{q}\left(i\gamma^{\mu} \partial_{\mu} -m_{0}\right)q
+\frac{G}{2} \left[(\bar{q} q)^{2}+(\bar{q} i\gamma_{5} \vec{\tau} q)^{2}\right],
\label{njl}
\end{equation}
where $m_{0}$ is the small current quark mass. 
At vanishing $m_{0}$ this NJL Lagrangian is invariant under chiral 
$ SU(2)_L \times  SU(2)_R $ transformations.  
The coupling constant $G$ has dimension $($energy$)^{-2}$, 
which makes the theory non-renormalizable. Therefore, a 
3-momentum cutoff $\Lambda$ is introduced to regularize divergent 
integrals. It defines an upper energy limit for this effective theory. 
Free parameters of the model are fixed to reproduce correctly the 
vacuum values of the pion decay constant (93 MeV), pion mass (138 MeV), and the constituent 
quark mass (337 MeV). Below we use the following parameters \cite{asakawa}:  
$G= 5.496$ GeV$^{-2}$,
$m_0 = 5.5$ MeV, and $\Lambda =631$ MeV. 
With these parameters the chiral transition occurs at the  
temperature $T\approx 190$ MeV \cite{klevanskyr} (for $\mu=0$) which is 
significantly higher than in the sigma model.

The partition function for the NJL model reads:
\begin{equation}
{\cal Z}={\rm Tr}\exp\left[-\left(\hat{\cal H}-\mu \hat{\cal N}\right)
\right]
= \int{\cal D}\bar{q}\, {\cal D}q\; \exp\left[\int_x\left({\cal L}+\mu 
\bar{q} \gamma^0q\right) \right]\,\, .
\end{equation}
In the mean-field approximation the Lagrangian (\ref{njl}) is represented 
in a linearized form \cite{weise,dhrwg}:
\begin{equation}
{\cal L}=\bar{q}(i\gamma^{\mu} \partial_{\mu} -m_{0})q
+ G{\langle}\bar{q}q{\rangle}(\bar{q}q)-\frac{G}{2}{\langle} \bar{q}q{\rangle}^2\,\, ,
\label{mfa}
\end{equation}
such that the partition function becomes
\begin{equation}
{\cal Z}=\exp\left[-\frac{V}{T}\frac{G\langle \overline{q}q\rangle^2}{2}\right ]
{\rm det}_p\left[(p_\mu\gamma^\mu + \mu\gamma^0 -M)/T\right],
\end{equation}
where the constituent quark mass is determined from the gap equation
\begin{equation}
M=m_0-G{\langle} \bar{q}q{\rangle}\,\, .
\label{njlmass}
\end{equation}
The right-hand side of this equation involves
the scalar density
\begin{equation}
\rho_s={\langle}\bar{q}q{\rangle}=M\nu_{q}\int_{p<\Lambda} \frac{d^3{\bf p}}{(2\pi)^3}
\; \frac{1}{E}[n_{q}(T,\mu)+n_{\bar{q}}(T,\mu)-1]\,\, ,
\label{expec}
\end{equation}
where $n_q$ and $n_{\bar{q}}$ are the valence quark and antiquark occupation
numbers defined in eq.\ (\ref{occunum}).
Here the last term in brackets gives the contribution from the 
Dirac sea (which corresponds to the vacuum part of the sigma model) 
and cannot be neglected. The rest comes from valence quarks 
and antiquarks similar to the sigma model (compare with eq.\ 
(\ref{scalardensity})).

From eq.\ (\ref{mfa}), the grand canonical potential for the NJL model 
can be written as:
\begin{equation}
\Omega = \frac{(M-m_0)^2}{2G}-
\nu_q\int_{p<\Lambda}\frac{d^3{\bf p}}{(2\pi)^3} \left\{ E +
T\ln\left[1+\exp\left(-\frac{E+\mu}{T}\right)\right]
+T\ln\left[1+\exp\left(-\frac{E-\mu}{T}\right) \right] \right\}.
\label{njlom}
\end{equation}
The minimization of $\Omega$ with respect to $M$ gives the gap equation
 (\ref{njlmass}).
The expression (\ref{njlom}) is formally identical with eq. (\ref{pqq}) derived
for the sigma model, but now the first term in curly brackets,
 coming from the 
Dirac sea, is treated explicitly after introducing the cut-off
momentum $\Lambda$. One can calculate  this
vacuum contribution explicitly,
\begin{equation}
\Omega_{vac}=-\nu_q\int_{p<\Lambda}\frac{d^3{\bf p}}{(2\pi)^3}\sqrt{p^2+M^2}=
-\frac{\nu_q\Lambda^4}{8\pi^2}\left [\sqrt{1+z^2}\left (1+\frac{z^2}{2} \right )-\frac{z^4}{2}\ln\frac{\sqrt{1+z^2}+1}{z} \right ],
\end{equation}
where $z=M/\Lambda$. Expanding this expression in powers of $z$ one can
find contributions to $\Omega$ of order $M^2$, $M^4$,.. . As mentioned 
above, in the sigma model the 
vacuum terms are partly absorbed in the coefficients of the effective
potential $U(\sigma,\vec{\pi})$. However, the logarithmic term
$M^4\ln\frac{\Lambda}{M}$ cannot be removed in this way.
Therefore, the NJL model has additional nonlinear terms in the vacuum
energy which are responsible for the differences in the thermodynamic
properties of the two models.

The sigma and pion masses are not as straightforward to obtain as in the 
linear sigma model because in the NJL model they are not 
represented as dynamical fields. In this model mesons
are described as collective $q\bar{q}$ excitations.
Their masses can be obtained from the poles
of the quark-antiquark scattering amplitude
 which can be 
computed, for instance, in Random Phase Approximation (RPA) \cite{klevanskyr}.
In this way, one can derive the following equations for the sigma and pion 
masses,
\begin{equation}
0=\frac{m_0}{M}+(M^2_{\sigma}-4M^2) G \,I(M,M_{\sigma})\,\,,
\end{equation}
\begin{equation}
0=\frac{m_0}{M}+M^2_{\pi} G \, I(M,M_{\pi})\,\,.
\end{equation}
The function $I(x,y)$ is the quark-antiquark propagator defined as:
\begin{equation}
I(x,y)=\frac{\nu_{q}}{2\pi^2}{\cal P}\int_{p<\Lambda} dp\, p^2\;
\frac{1}{E}[1-n_{q}-n_{\bar{q}}]\frac{1}{E^2-\frac{1}{4}y^2},
\end{equation}
where $E=\sqrt{x^2+p^2}$, and the occupation numbers $n_q,\, n_{\bar{q}}$
are as defined in (\ref{occupationnumbers}). 
In this integral ${\cal P}$ means principal value. 

\section{Numerical Results}

In the case of the linear sigma model everything is determined when the 
gap equation
(\ref{smmass}) is solved in the $(T,\mu)$ plane, whereas in the NJL model
the meson masses have to be solved for as well.
Below we present results of our numerical calculations (see also Appendix).

\subsection{Phase diagrams}
\begin{figure}[htp]
\centerline{\hbox{\epsfig{figure=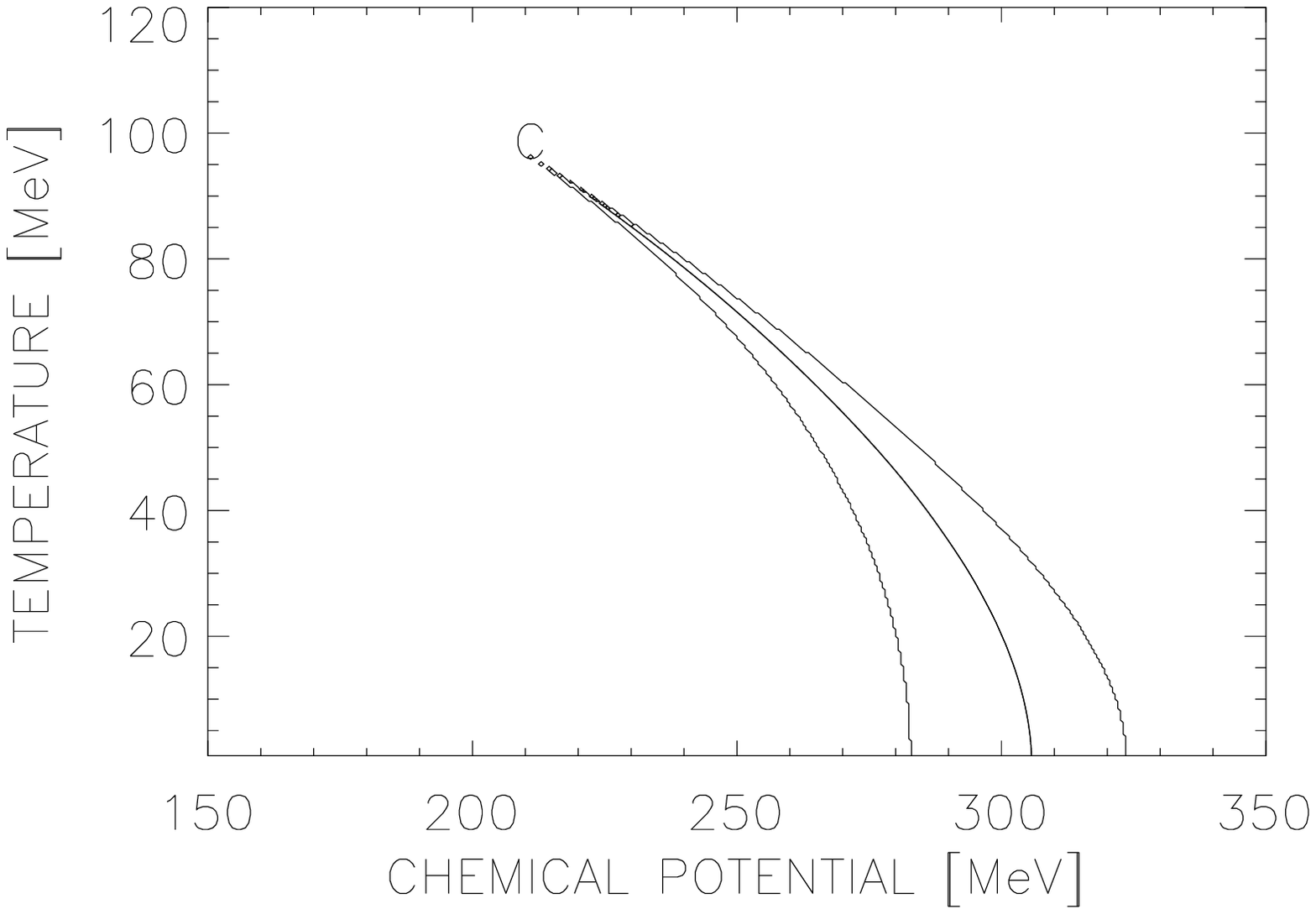,height=6.5cm} \hspace{-1.0cm}
\epsfig{figure=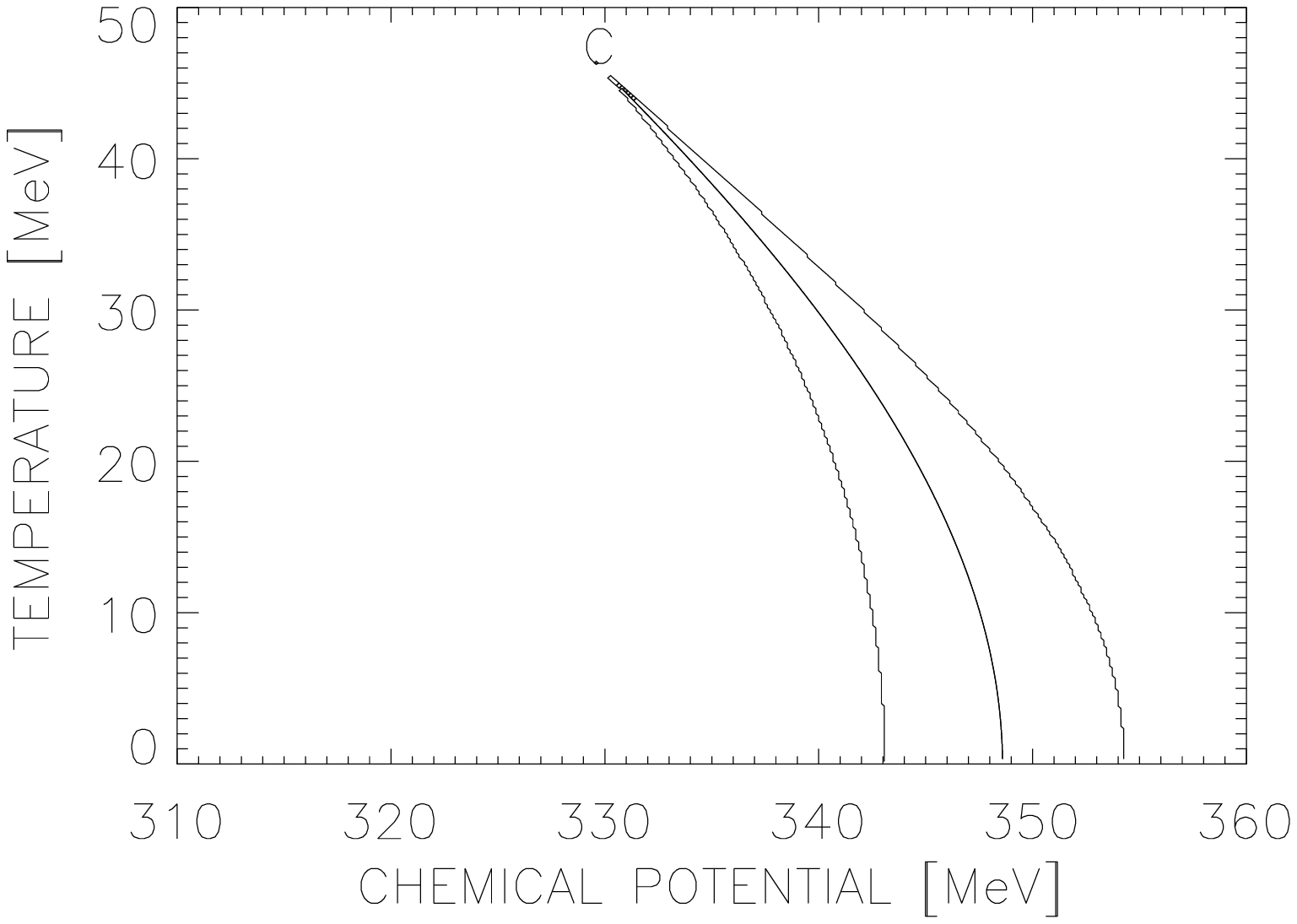,height=6.5cm}}}
\caption{The phase diagrams for the sigma model (left) and the 
NJL model (right) in the ($\mu,T$) plane. The middle curve is the critical line
and the outer lines are the lower and upper spinodal lines. C is the
critical point.}
\label{ptdia}
\end{figure}

We start this section with presenting in Fig. \ref{ptdia} the resulting
phase diagrams in the $(T,\mu)$ plane calculated for the two models.
The middle line corresponds to the states  where the 
 two phases co-exist in the first order phase transition.
Along this line the thermodynamical potential $\Omega$ has two minima of equal
depth separated by a potential barrier which height grows towards
lower temperatures. At the critical point C the barrier
disappears and the transition is of second order.
The other lines in Fig. \ref{ptdia}
 are spinodal lines which constrain the regions of spinodal instability
where $\left ( \frac{\partial n_B}{\partial \mu} \right )_T <0$.
Information about the timescales of this instability can be obtained 
from dynamical simulations \cite{ms,oa}.

\begin{figure}[htp]
\centerline{\hbox{\epsfig{figure=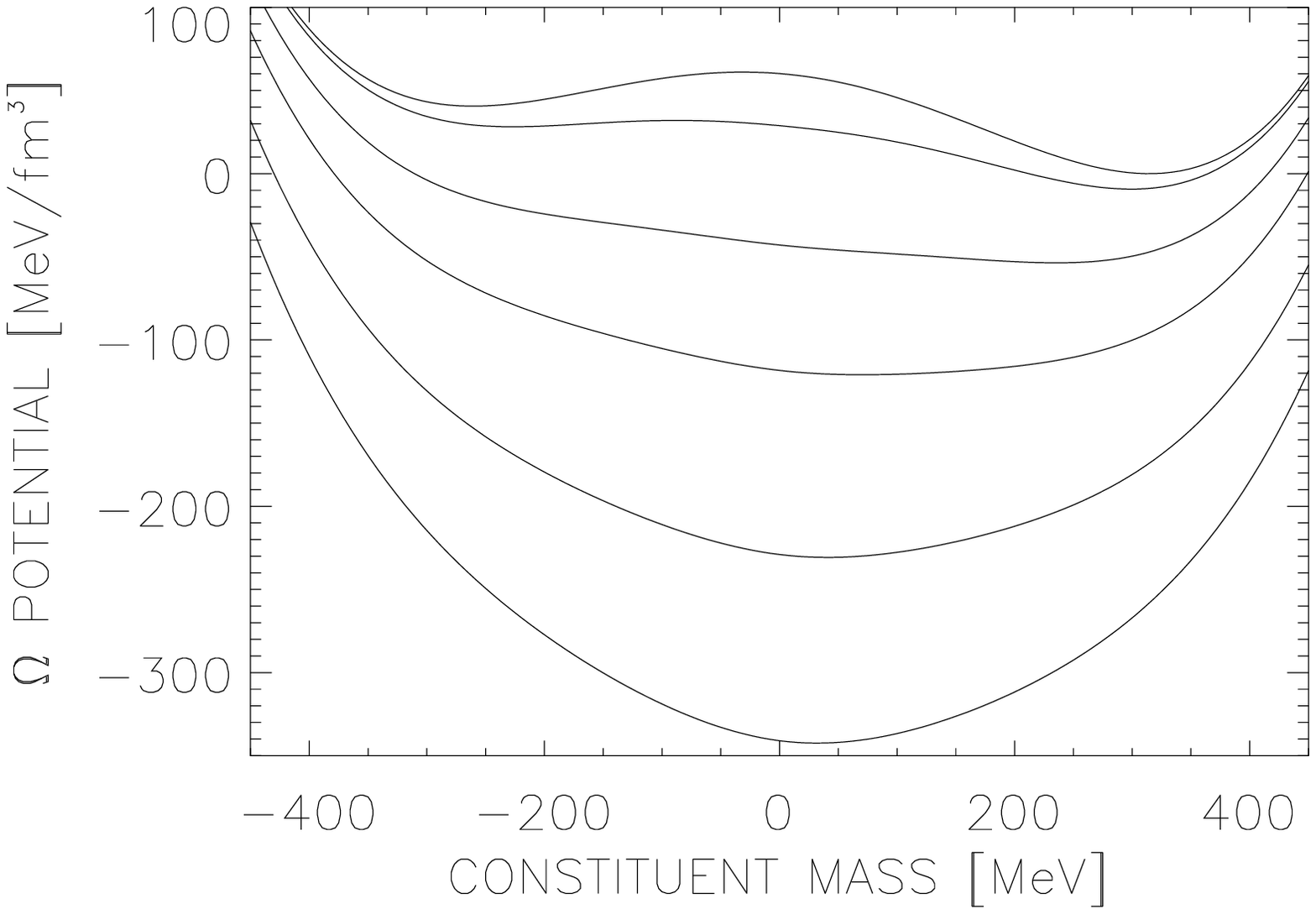,height=6.5cm} \hspace{-1.0cm}
\epsfig{figure=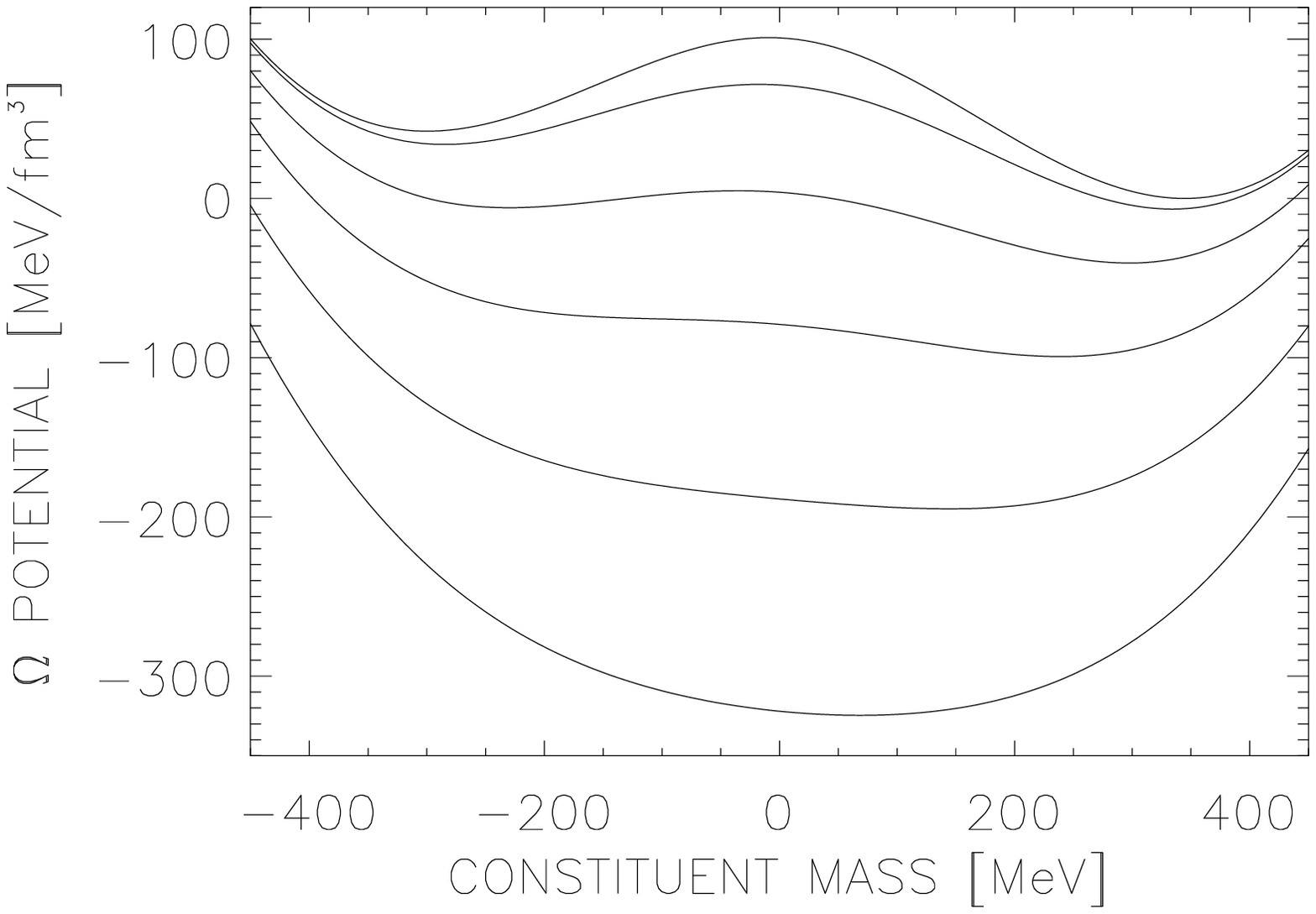,height=6.5cm}}}
\caption{The thermodynamical potentials $\Omega$ for the sigma model
(left) and the NJL model (right). 
For both models $\mu=0$. The levels correspond to 
(starting from the top):
$T=[0,100,135,155,175,190]$ MeV for the sigma model and 
$T=[0,100,140,170,200,230]$ MeV for the NJL model. }
\label{pott}
\end{figure}

It is instructive to plot the thermodynamic potential 
as a function
of the order parameter for various values of $T$ at $\mu=0$, and
for various values of $\mu$ at $T=0$.
The first case is shown in Fig. \ref{pott}, where the left panel is for
the sigma model and the right panel for the NJL model.
One clearly sees the smooth crossover of the symmetry breaking pattern in 
both cases. Note that the effective bag constant (the energy difference
between the global minimum and the local maximum of the
potential in vacuum) is about 
$100$ MeV/fm$^3$ in the NJL model,
whereas in the sigma model it is significantly smaller,  
$\simeq 60$ MeV/fm$^3$. 
To a large extent this difference is responsible for the
difference in the temperatures corresponding to the crossover
transition: about 140 MeV in the sigma model and about 180-190 MeV
in the NJL model.

\begin{figure}[htp]
\centerline{\hbox{\epsfig{figure=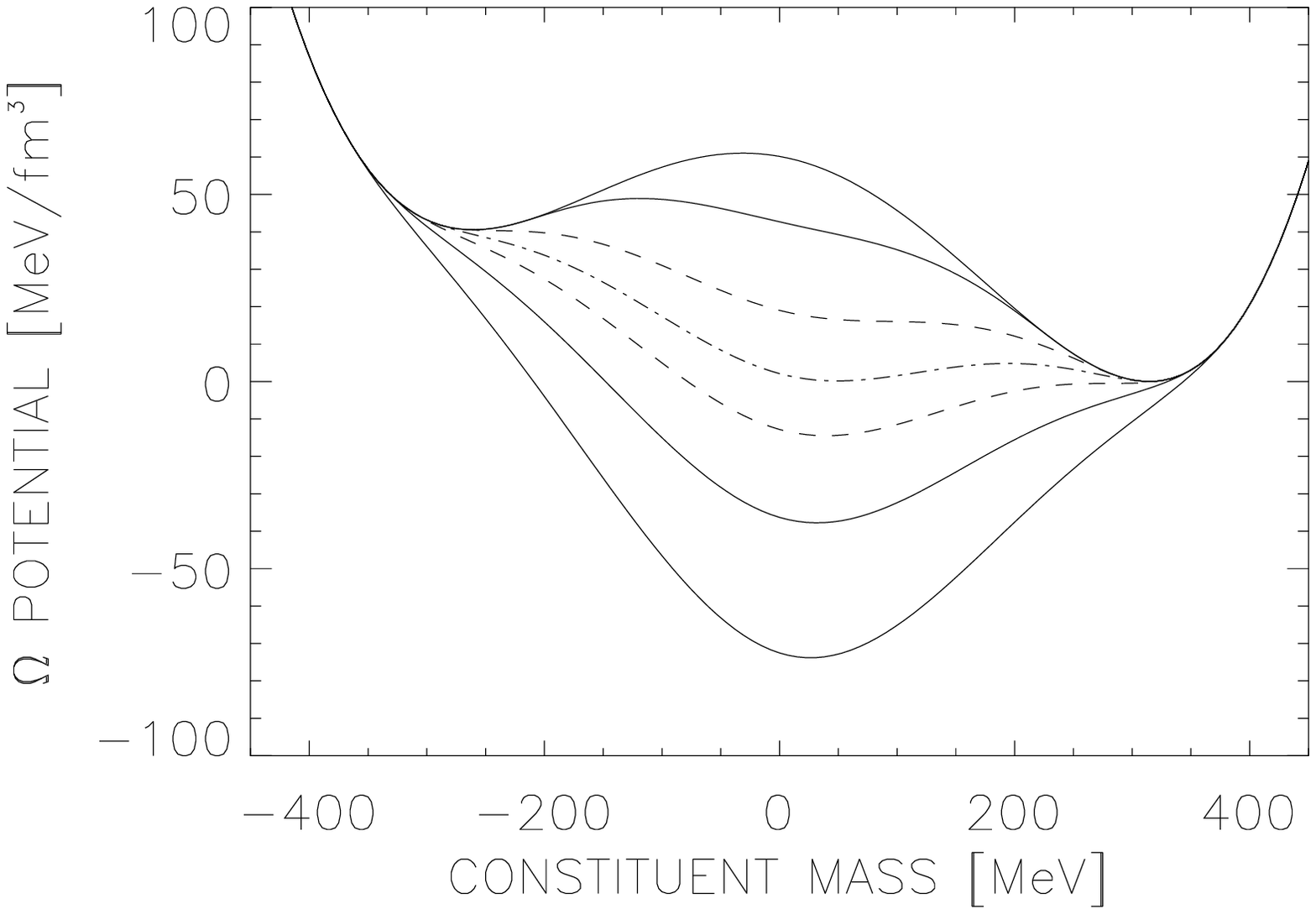,height=6.5cm} \hspace{-1.0cm}
\epsfig{figure=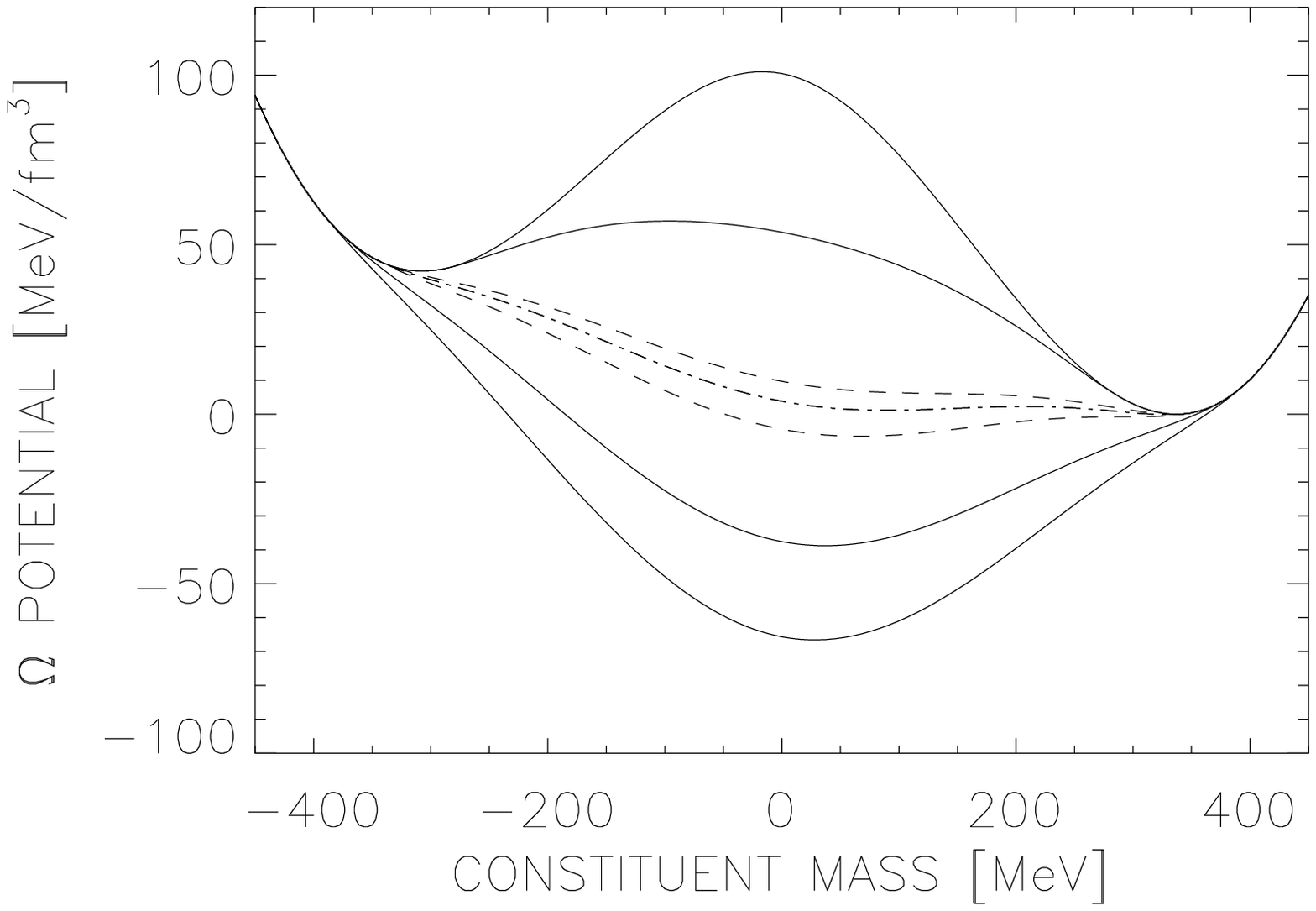,height=6.5cm}}}
\caption{The thermodynamical potentials $\Omega$ for the sigma model
(left) and the NJL model (right). 
For both models $T=0$. The levels correspond to (starting from the top):
$\mu=[0,225,279,306,322,345,375]$ MeV in the sigma model,
and 
$\mu=[0,288,343,348,35,378,396]$ MeV for the NJL model.}
\label{potmu}
\end{figure}

\begin{figure}[htp]
\centerline{\hbox{\epsfig{figure=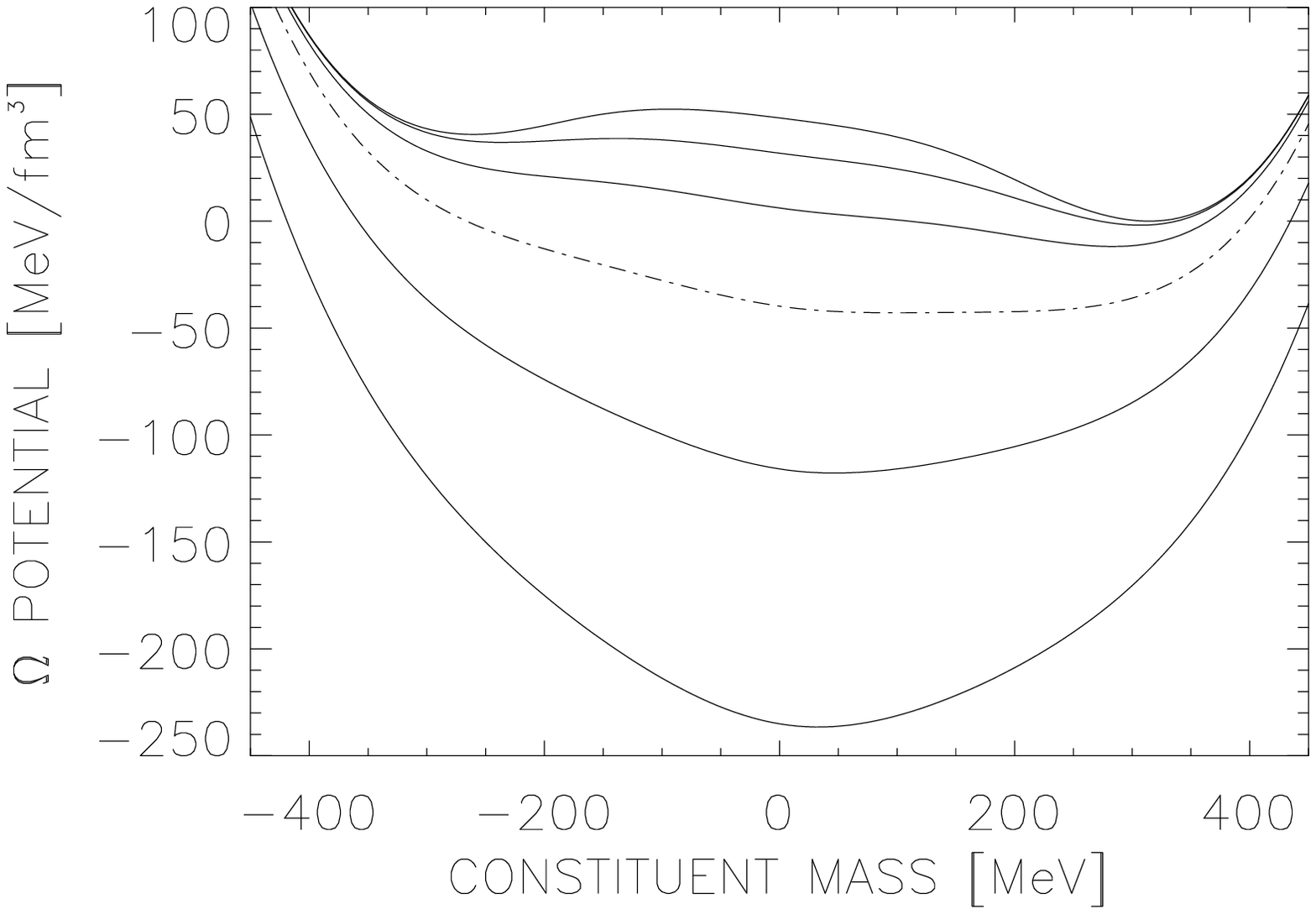,height=6.5cm} \hspace{-1.0cm}
\epsfig{figure=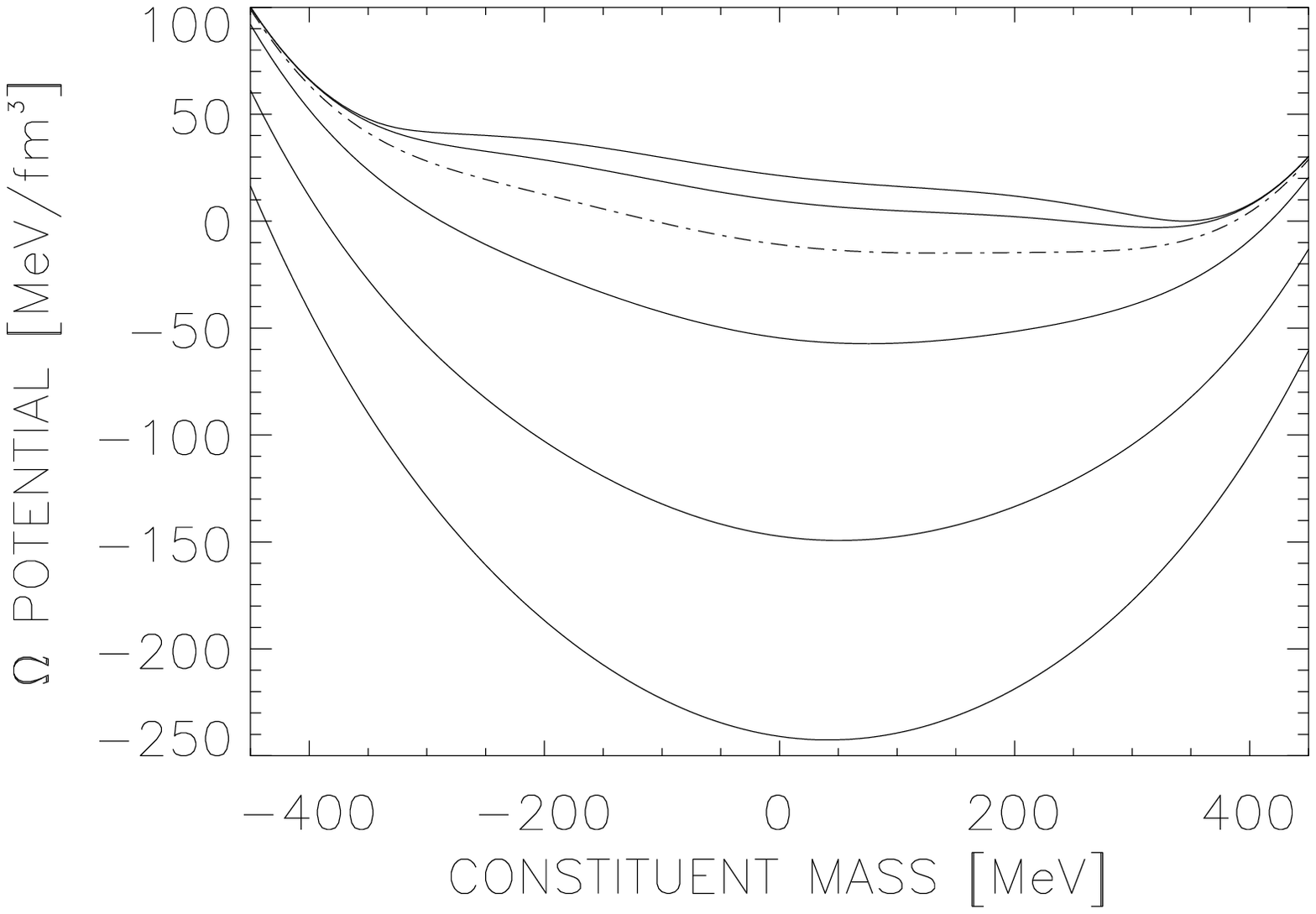,height=6.5cm}}}
\caption{The thermodynamical potentials $\Omega$ for the sigma model
(left) and the NJL model (right). 
For the sigma model case $\mu$ is fixed to 207 MeV and the levels 
correspond to (starting from the top):
$T=[0,50,75,100,125,150]$ MeV.
For the NJL model case $\mu$ is
fixed to 332 MeV and the levels correspond to (starting from the top):
$T=[0,28,46,70,105,133]$ MeV.
} 
\label{potcrit}
\end{figure}

In Fig. \ref{potmu} the same plot is shown for $T=0$ and a nonzero $\mu$.
Here, one clearly observes the pattern characteristic for 
a first order phase transition: two minima corresponding 
to phases of restored and broken symmetry separated by a potential barrier. 
The barrier height is larger in the 
sigma model than in the NJL model, thus, indicating a weaker first order
phase transition in the NJL model. It now follows that somewhere in between
these two extremes, for some $\mu_c$ and $T_c$, there exists a 
second order phase transition (the critical point). 
Indeed, this point is found  and shown in Fig. \ref{ptdia}. The corresponding
values are $(T_c,\mu_c)\simeq (99,207)$ MeV in the sigma model, 
and $(T_c,\mu_c)\simeq(46,332)$ MeV in the NJL model. The behavior
of the thermodynamic potential at $\mu=\mu_c$ and various $T$ is shown
in Fig. \ref{potcrit}. One can see that the potential has only
one minimum which is flattest at the critical point.

\subsection{Effective Masses}
\begin{figure}[htp]
\centerline{\hbox{\epsfig{figure=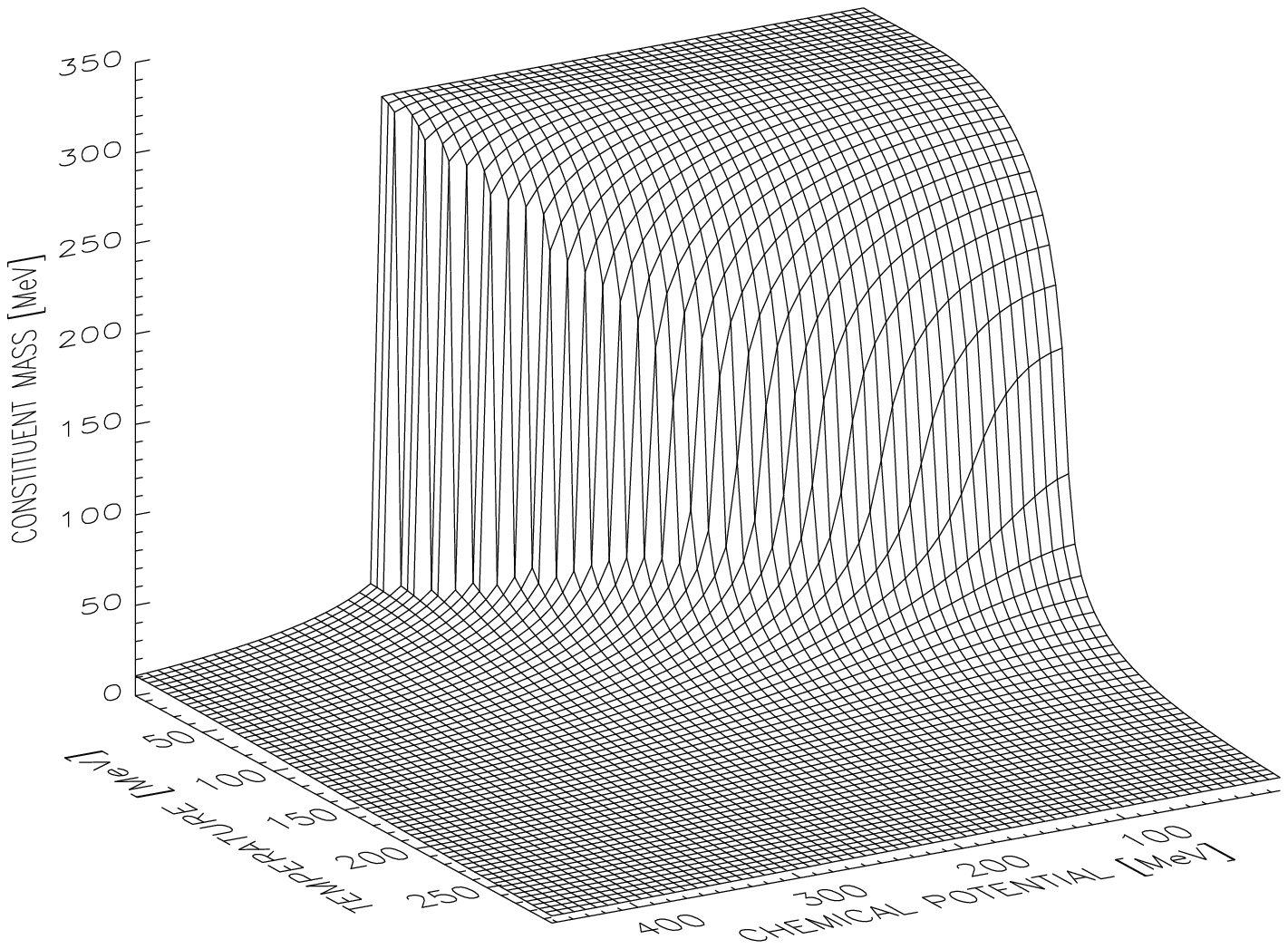,height=6.5cm} \hspace{-1.0cm}
\epsfig{figure=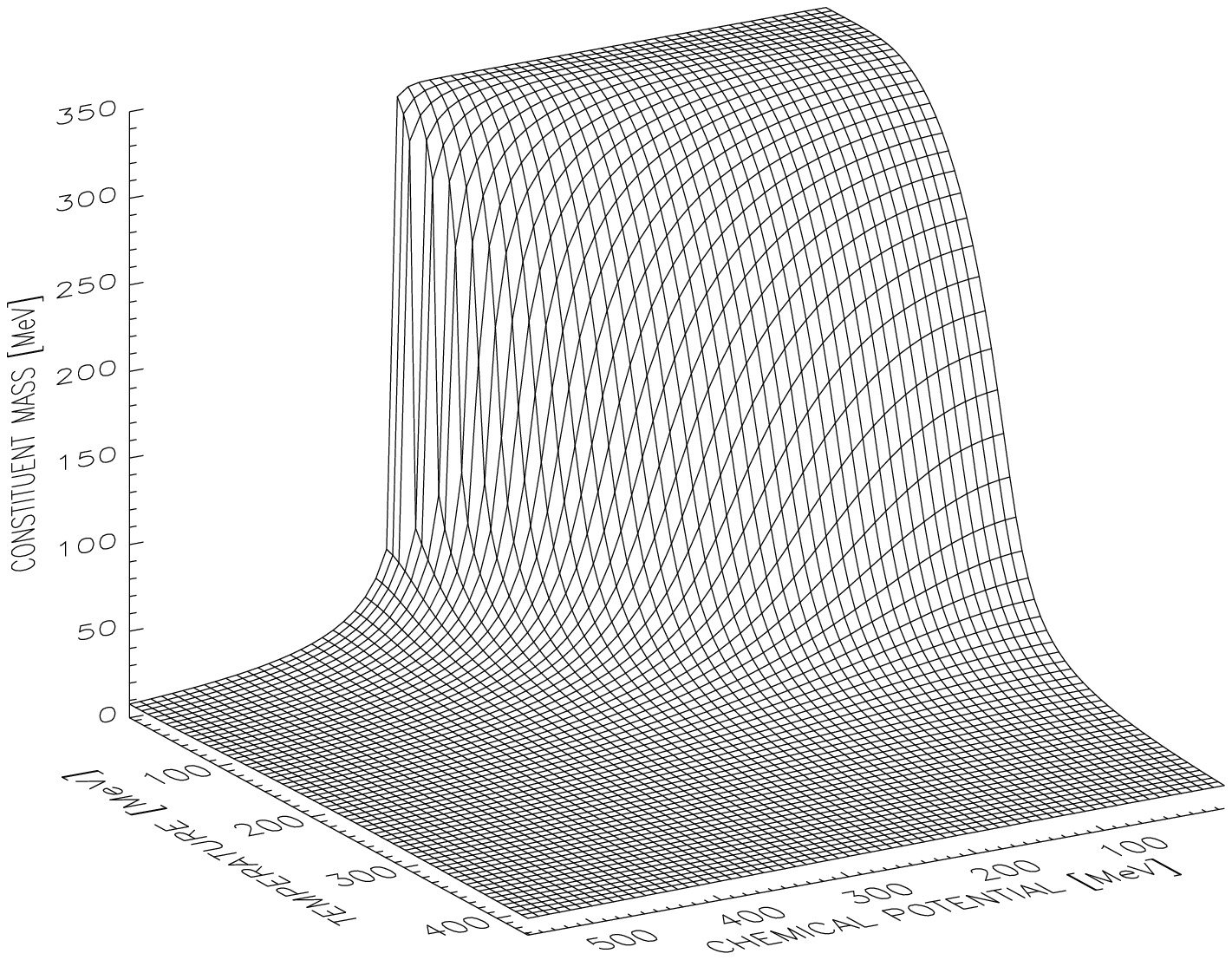,height=6.5cm}}}
\caption{The constituent quark (antiquark) mass in
the sigma model (left) and the NJL model (right) as function
of $\mu$ and $T$.
}  
\label{qmassfig}
\end{figure}

Now let us consider the model predictions for the effective
masses. The constituent quark mass is shown in Fig. \ref{qmassfig} as function of $T$ and $\mu$. These plots, of course, 
show the same phase structure as discussed above. 
At $\mu=0$ in both models the quark mass
 falls smoothly from the respective vacuum value and approaches zero 
as $T$ goes to infinity. One could define a crossover
 temperature as corresponding to a steepest descent region in the variation
of $M$. This again gives a temperature of 
 about $140-150$ MeV for the sigma model and about $180-190$ MeV for
the NJL model. At $T=0$ and nonzero $\mu$ the constituent quark 
mass shows a discontinuous behavior  reflecting a first order chiral
transition.

\begin{figure}[htp]
\centerline{\hbox{\epsfig{figure=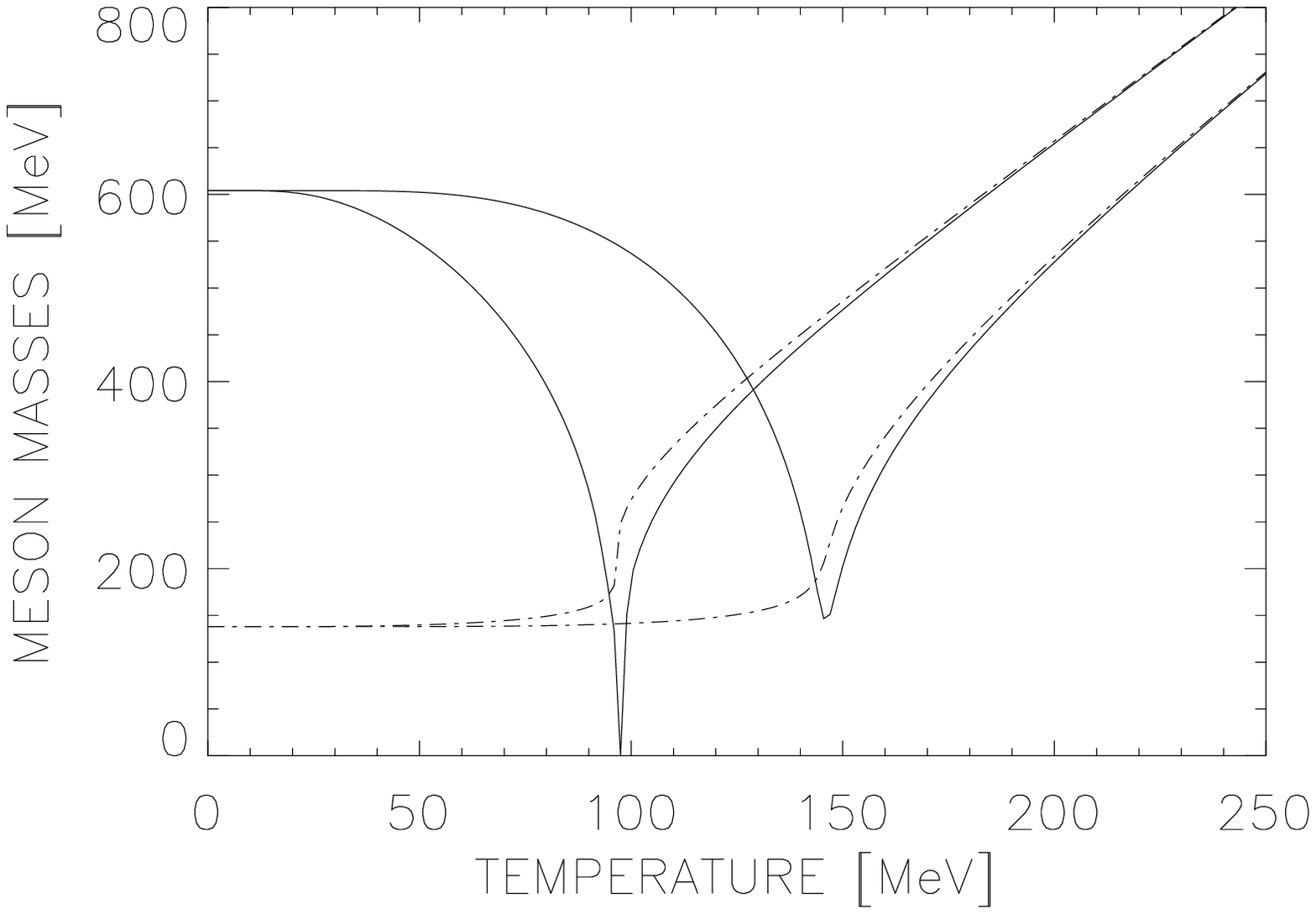,height=6.5cm} \hspace{-1.0cm}
\epsfig{figure=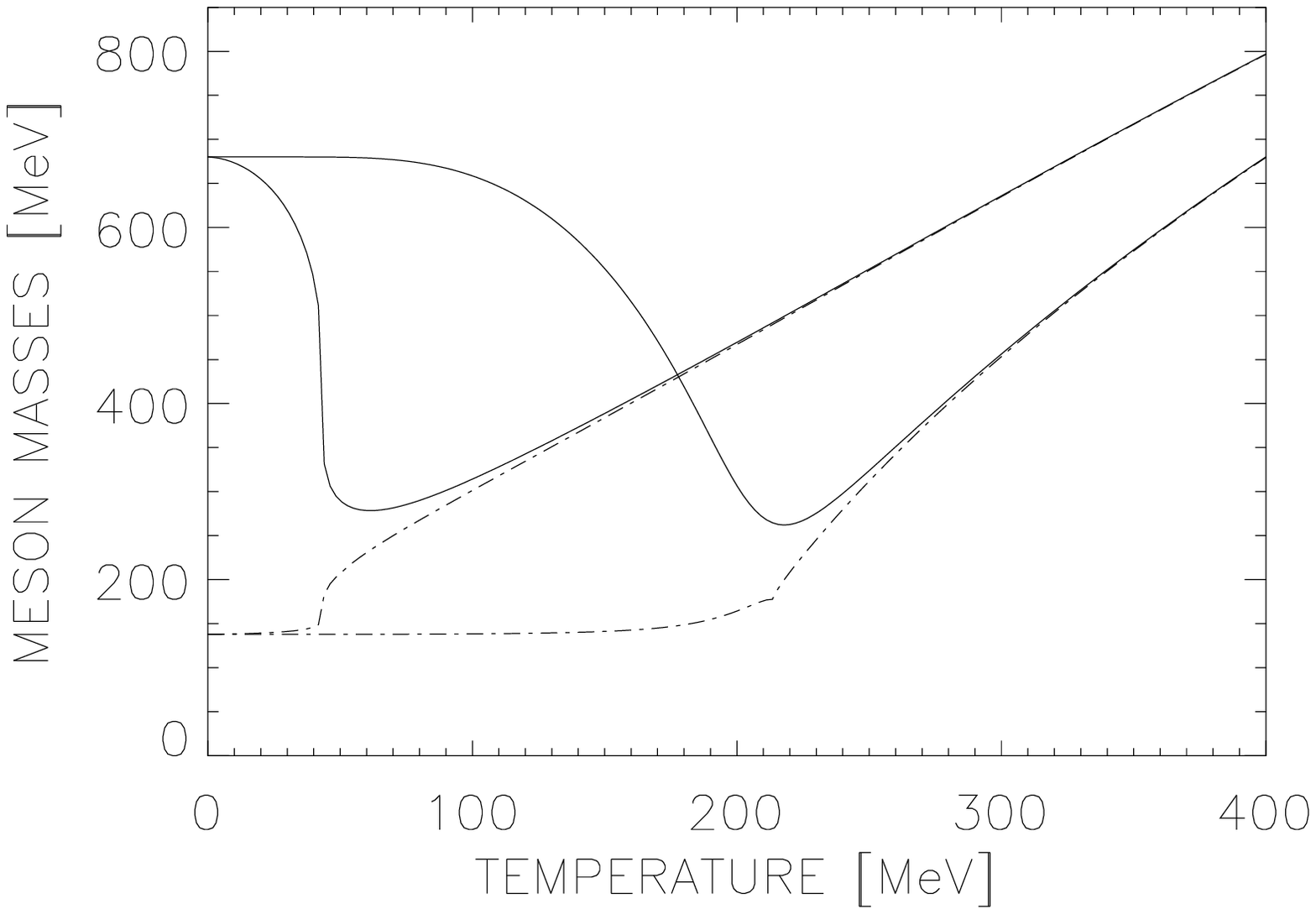,height=6.5cm}}}
\caption{
The sigma mass (solid line) and pion mass (dashed line) in the
sigma model (left) and NJL model (right) as functions
of temperature for $\mu=0$ (right pair) and for $\mu=\mu_{c}$ (left pair).
} 
\label{mesmasst}
\end{figure}

\begin{figure}[htp]
\centerline{\hbox{\epsfig{figure=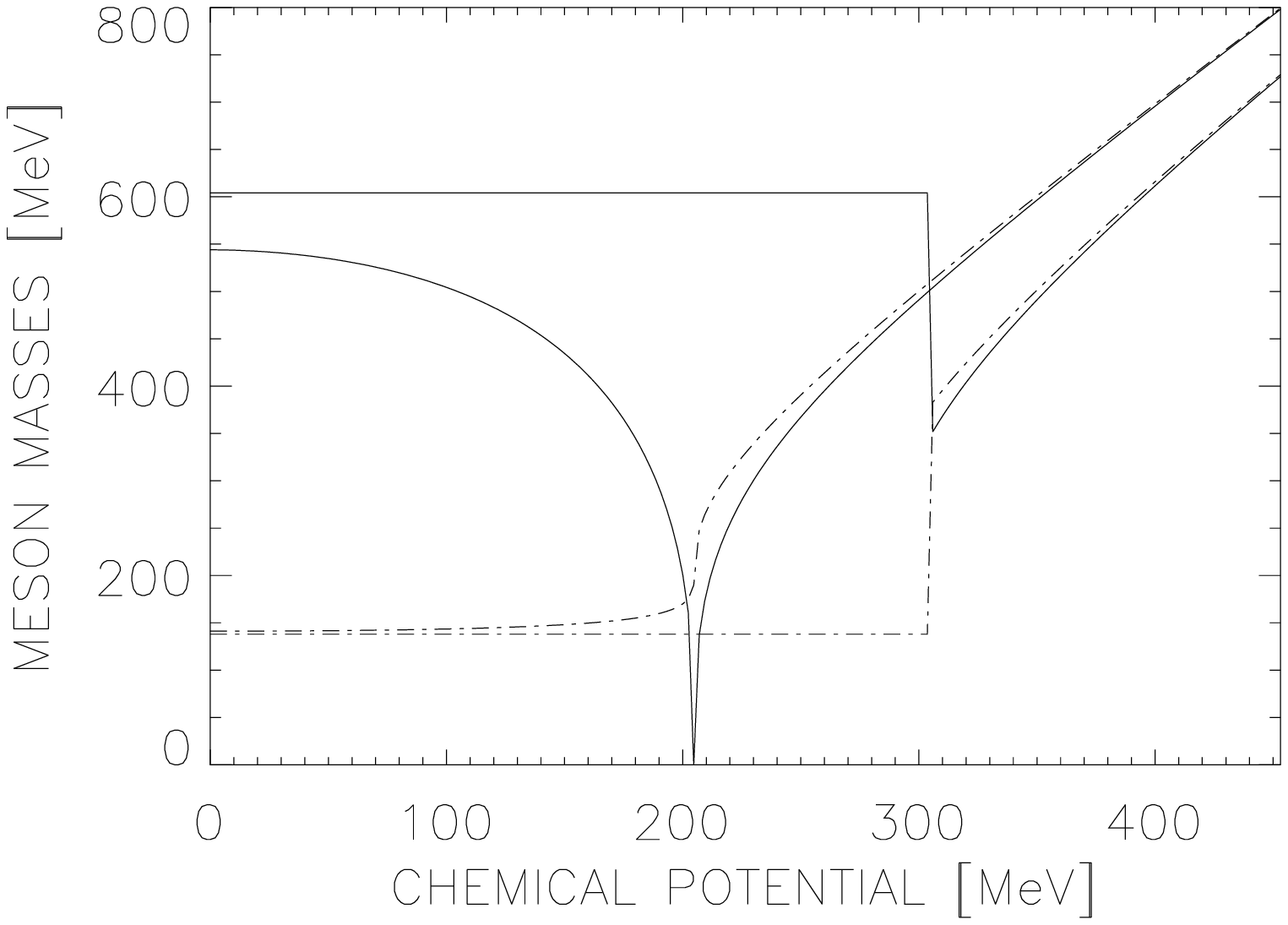,height=6.5cm} \hspace{-1.0cm}
\epsfig{figure=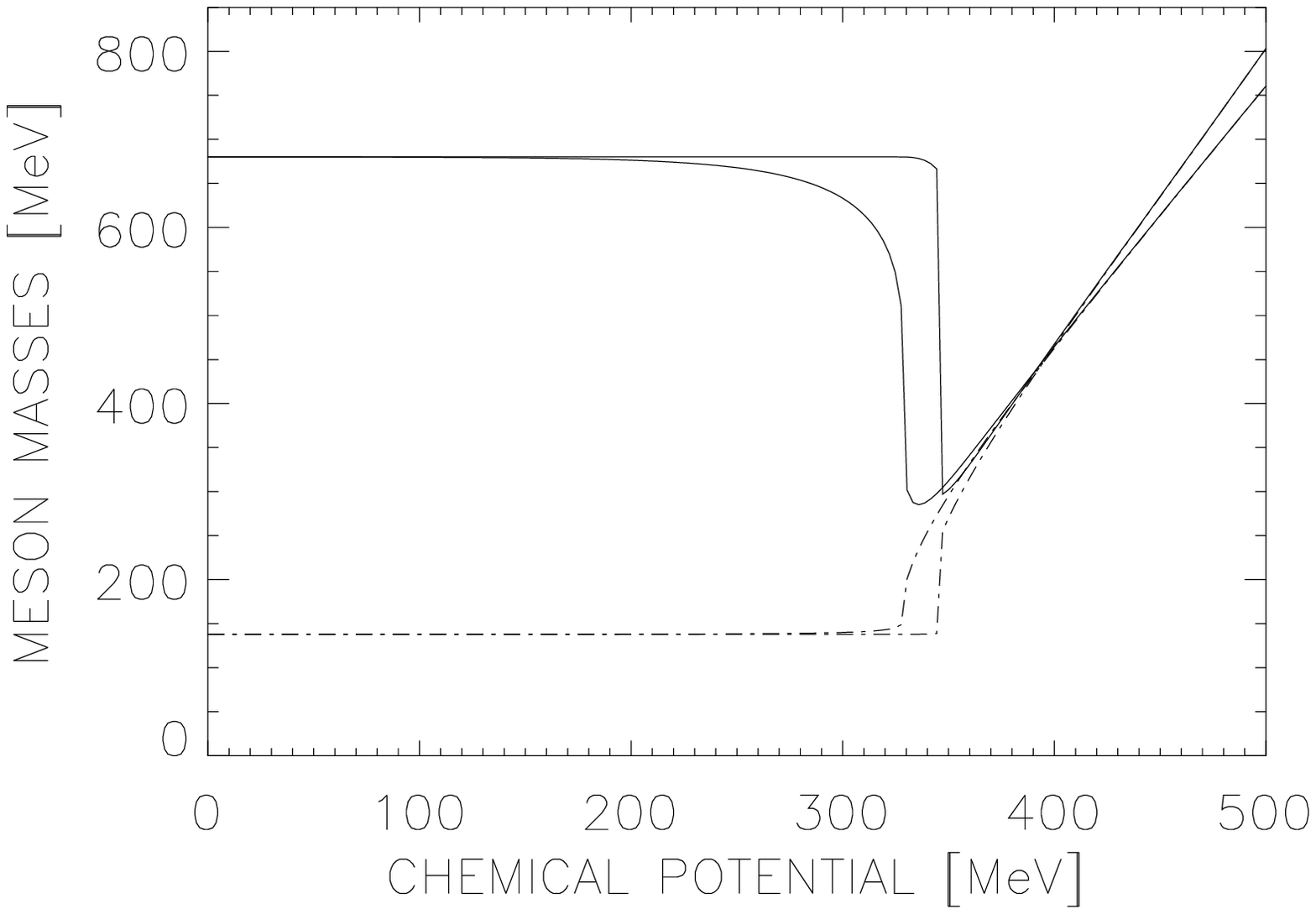,height=6.5cm}}}
\caption{The sigma mass (solid line) and pion mass (dashed line) in the
sigma model (left) and NJL model (right) as functions
of chemical potential for $T=0$ (right pair) and for $T=T_{c}$ (left pair).
} 
\label{mesmassmu}
\end{figure}

The sigma and pion masses for various $T$ and $\mu$ are shown in Figs. 
\ref{mesmasst},\ref{mesmassmu}.
In both models the sigma mass first decreases smoothly, 
then rebounds and grows again at high $T$. 
The pion mass does not change much at temperatures below $T_c$
but then increases rapidly, approaching the sigma
mass and signaling the restoration of chiral symmetry.
As $T$ goes to infinity the masses 
grow linearly with $T$. The $\mu=\mu_c$ case is especially 
interesting in the sigma model.
Since the sigma field is the order parameter of
the chiral phase transition, its mass must vanish at the critical
point for a second
order phase transition. This means that $\Omega$ has zero curvature at
this point. 
It is, however, not clear what the sigma mass should be at the critical 
point in the NJL model where the quark condensate $\langle \bar{q}q\rangle$
is the order parameter.
 Fig. \ref{mesmasst} indeed shows that 
exactly at the critical point the sigma mass
is zero in the sigma model. This is not the case in the NJL model, 
at least within the RPA used here.  

In Fig. \ref{mesmassmu} the masses are plotted as function of $\mu$ for $T=0$ and
$T=T_c$. For $T=0$ one clearly sees discontinuities in the 
behavior of the masses characteristic for the
first order phase transition.

An interesting point is that, in the linear sigma model,
 there is no stable phase with heavy
quarks for $T=0$, i.e., the quark mass assumes its vacuum value
all the way up to the chiral transition, and then drops
to a small value in the phase where chiral symmetry is restored (see 
Fig. \ref{qmass}).
This behavior is related to the appearance of a bound state at zero pressure.
Within the linear sigma model this ``abnormal'' bound state was found
by Lee and Wick a long time ago \cite{leewick}.
Recently, it was shown in ref. \cite{mishusat} that a similar bound state
appears also in the NJL model.
This behavior, however, depends on the value of the coupling constant $g$
or $G$. For our choice of $g$ and $G$, this state exists in the linear
sigma model, but not in the NJL model, where there is a stable
phase of heavy quarks at $T=0$, cf. Figs. \ref{qmassfig} and \ref{mesmassmu}.
 In general, if the coupling constant is sufficiently large, 
then the attractive force between the constituent quarks
becomes large enough to counterbalance the Fermi pressure, thus giving rise to
a bound state. To demonstrate this 
we have varied the coupling constant $g$ for the sigma model within reasonable
limits.

%%%%%%%%%%%%

\begin{figure}[htp]
\centerline{\hbox{\epsfig{figure=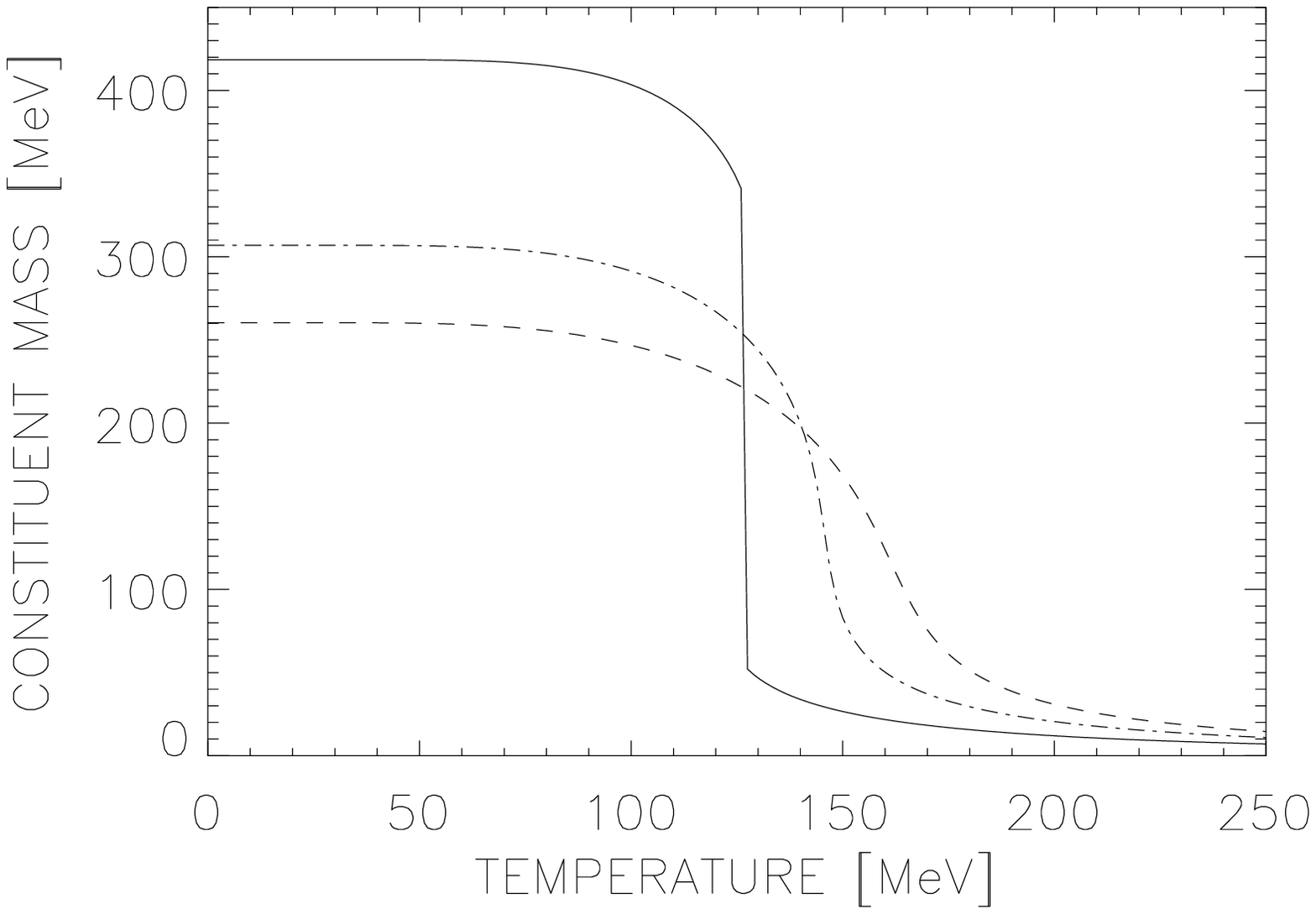,height=6.5cm} \hspace{-1.0cm}
\epsfig{figure=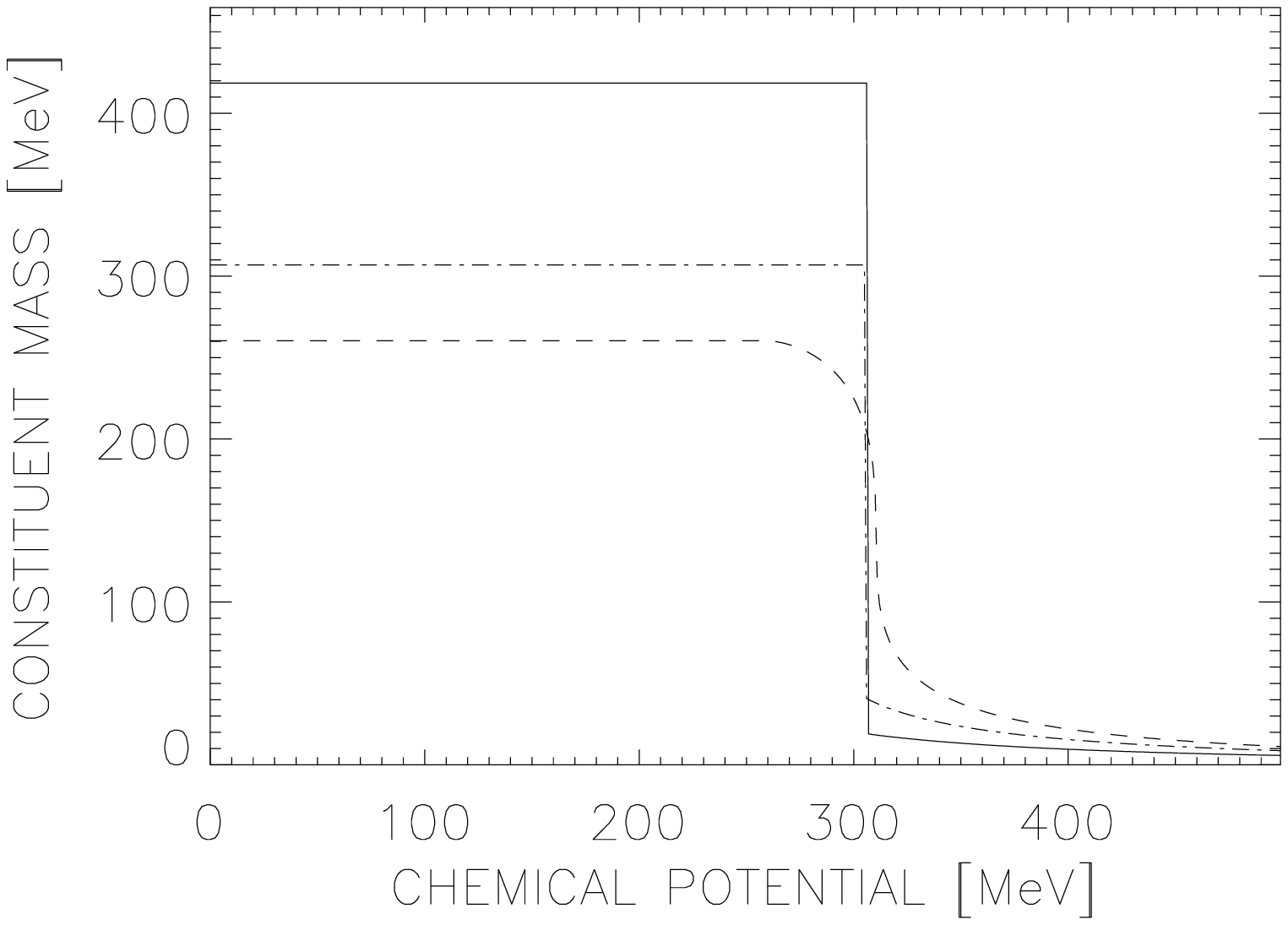,height=6.5cm}}}
\caption{The constituent quark (antiquark) mass as function
of temperature for zero chemical potential (left) and as function
of chemical potential (right) for zero temperature in the sigma model. 
The solid line represents
the mass for $g=4.5$, the dashed-dotted line for $g=3.3$ and
 the dashed line for $g=2.8$.
} 
\label{coup}
\end{figure}

The results for the quark mass are shown in Fig. \ref{coup}.
It is seen that, indeed, one can change the smooth crossover for $\mu=0$ into
a first order transition by increasing the coupling constant (Fig. \ref{coup}
 left panel)
and change the first order transition in the case\ of $T=0$ into
a smooth crossover. In this way a heavy quark phase comes into existence as
the coupling constant is decreased and the bound state disappears
 (Fig.\ref{coup} right panel). An analogous investigation
for the NJL model leads to similar results.

\subsection{Adiabats}
\begin{figure}[htp]
\centerline{\hbox{\epsfig{figure=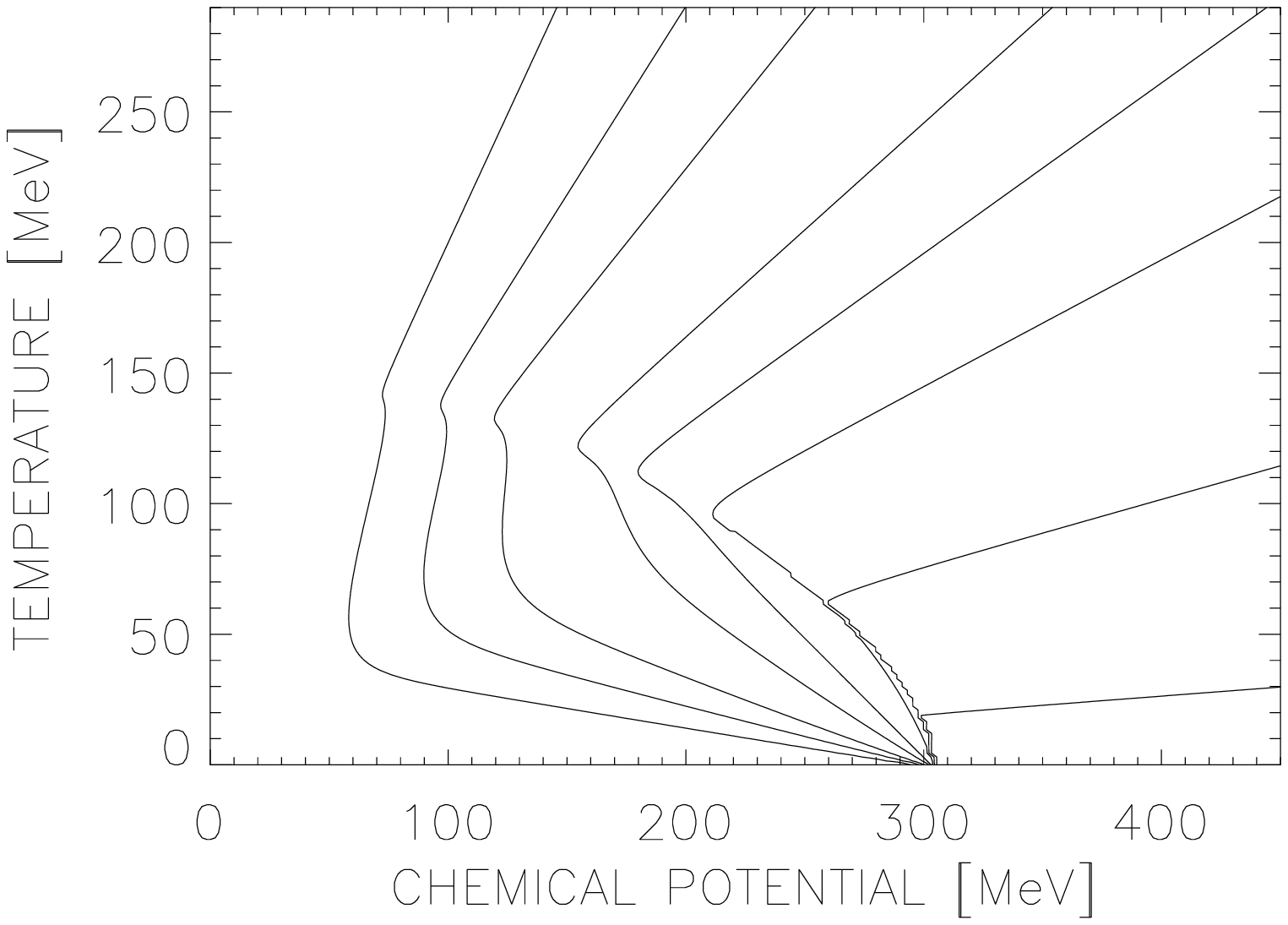,height=6.5cm} \hspace{-1.0cm}
\epsfig{figure=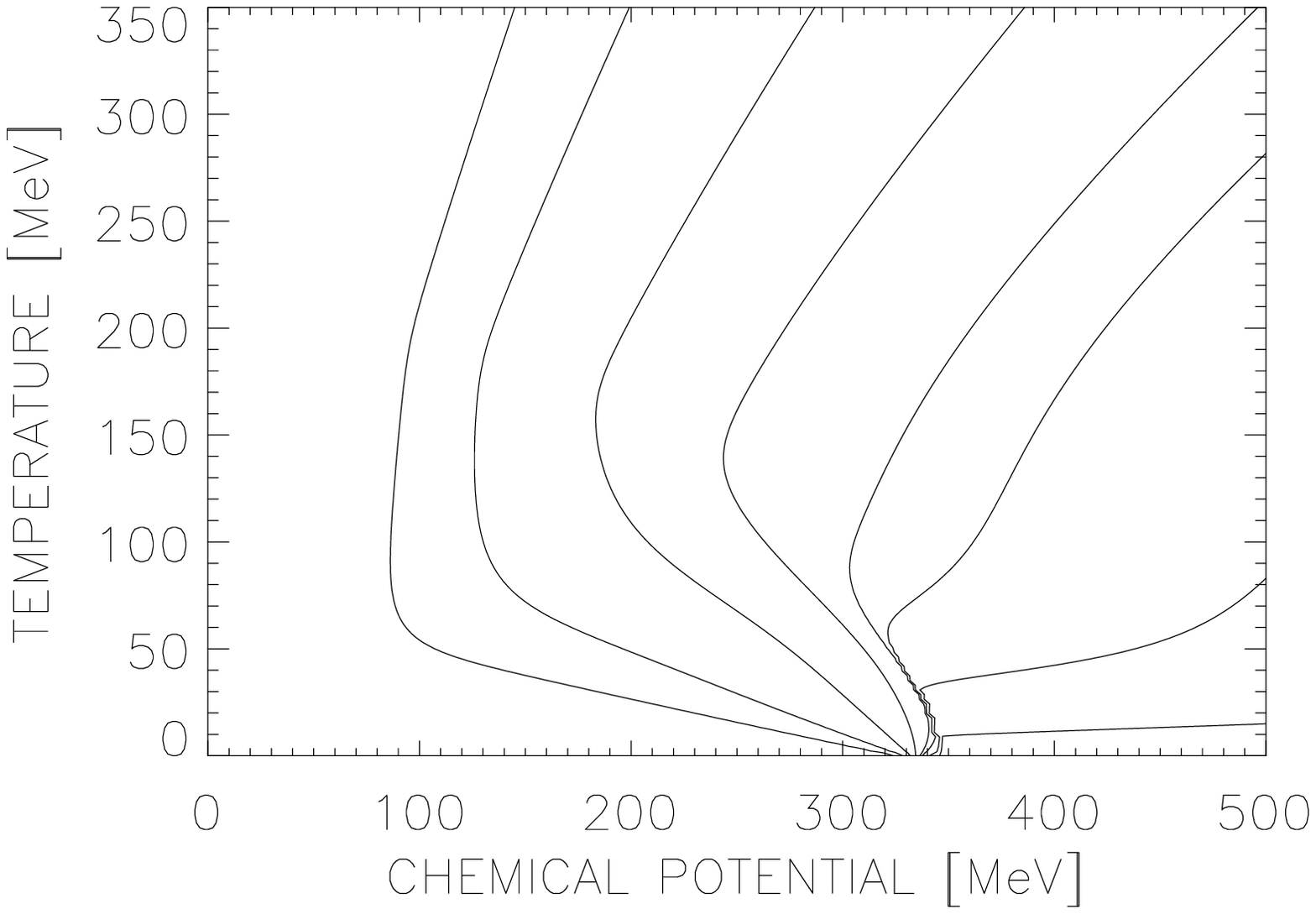,height=6.5cm}}}
\caption{The entropy per baryon number, $S/A$ for the
sigma model (left) and the NJL model (right).
In the sigma model the curves correspond to (from left) 
$S/A$=[28,21,17,13,11,9,6,2].
In the NJL model the curves correspond to (from left)
$S/A$=[22,16,11,8,6,5,3,1].
} 
\label{adiabats}
\end{figure}

Regarding  hydrodynamical simulations the entropy per baryon is an interesting
quantity. One can easily calculate it using standard thermodynamic relations,
\begin{equation}
\frac{S}{A}=3\frac{e+p-\mu n}{Tn},
\end{equation}
where $e,p,n$ are respectively the energy density, pressure and net 
density of
the quarks and antiquarks ($n=3n_B$).
By studying this quantity, one can check if there is a tendency towards convergence of the adiabats towards the critical 
point as was claimed in ref. \cite{SRS}.
If this was the case it would be easy to actually hit or go close
to this point in a hydrodynamical evolution.
Fig. \ref{adiabats} shows the contours of $S/A$ in the $(T,\mu)$ plane calculated in the
sigma model (left) and in the NJL model (right).
We actually observe a trend which is quite opposite to this expectation. It turns out that the adiabats turn away from the critical
point when they hit the first order transition line and bend 
towards the critical point only when they 
come from the smooth crossover region. 
This is explained as follows.
First, note that all adiabats terminate at zero temperature and
$\mu=M_{vac}$, i.e. the $(T,\mu)$ combination corresponding to the vacuum. 
The reason is that as $T\rightarrow 0$, also $S\rightarrow 0$
(by the third law of thermodynamics), therefore, for fixed
$S/A$ we have to require that $n\rightarrow 0$, which is fulfilled
when $\mu=M_{vac}$. For our choice of parameters, in the sigma model the
 point $(T,\mu)=(0,M_{vac})$
is also the endpoint of the
phase transition curve at $T=0$, since the phase transition connects
the vacuum directly with the phase of restored chiral symmetry,
 cf. Figs. \ref{qmassfig} and \ref{coup}. For the NJL model,
the endpoint of the phase transition curve is not identical
with $(0,M_{vac})$ but is rather close to it.
Therefore, also the adiabats which hit the phase transition curve
have to bend away from the critical point and approach the endpoint
of the phase transition line at $T=0$, i.e., $T$ decreases and $\mu$ increases.

This behavior is quite opposite to the case underlying the
claim in ref. \cite{SRS}, where the hadronization
of a large number of quark and gluon degrees of freedom into
relatively few pion degrees of freedom
leads to the release of latent heat and consequently to
a reheating (increase of $T$) through the phase transition. 
Remember, however, that in our case there is actually no change in the
number of degrees of freedom in the two phases, only in their 
respective masses. Consequently, there is no ``focusing''
effect in the linear sigma and NJL models.

%This behavior originates in the fact that during the phase 
%transition the number of degrees of freedom does not change, only the constitu%ent
%quark mass changes. Since the quark chemical potential must be larger
%than the constituent mass the adiabats bend towards larger chemical
%potentials in the broken phase. 
%This is why the temperature decreases while the chemical potential increases
%along the phase boundary.

\section{Conclusions}

We investigated the thermodynamics of the chiral phase transition within the linear sigma model coupled to quarks and 
the Nambu-Jona-Lasinio model. These models have similar vacuum properties
 but treat the contribution of the Dirac Sea differently. In the sigma model
this contribution is ``renormalized out'' while in the NJL model
it is included explicitly up to a  momentum cutoff $\Lambda$.
By comparing thermodynamic properties of these two models one can check the 
importance of these vacuum terms. In both models, we found for small bare quark masses a smooth crossover for nonzero temperature and 
zero chemical potential and a first order transition for zero temperature
and nonzero chemical potential. The first order phase transition 
line in the ($T,\mu$) plane ended in the expected critical point. It has been 
found that the $\sigma$ mass is zero at 
the critical point in the sigma model whereas in the NJL model it
always remains nonzero.
The phase transition in the sigma model turned out to be of the liquid-gas
type. This, however, depends on the coupling constant $g$ between the quarks and the chiral fields. From the comparison we conclude that the phase transition
pattern is generally weaker in the NJL model than the sigma model. Certainly,
it will be interesting to use both models in hydrodynamical
simulations in order to confirm or disconfirm possible observable
signatures of the phase transition discussed in the introduction. In particular, the 
sigma model which contains dynamical $\sigma$ and pion  fields, would be
suitable to study the long wavelength enhancement of the $\sigma$ field at 
the critical point.  Such simulations are in progress.  

\section*{Acknowledgements}
The authors thank J. Borg, L.P. Csernai, P. Ellis, A.D. Jackson,
 L.M. Satarov and A. Wynveen for useful discussions. 
O.S. thanks the Yale Relativistic Heavy Ion Group for kind hospitality and
support from grant no.\ DE-FG02-91ER-40609.
The work of \'A.M. is supported by the
US. Department of Energy under grant DE-FG02-87ER40328.
I.N.M. thanks the Humboldt
Foundation for financial support.
D.H.R thanks RIKEN, BNL, and the U.S. Dept. of Energy for providing
the facilities essential for the completion of this work.

\appendix
\section{}
In the chirally symmetric phase the constituent quark (antiquark) mass
$M$ is small, it goes to zero in the chiral limit. Therefore, it is 
instructive to evaluate the thermodynamic potential
$\Omega(T,\mu,M)$ for small $M$. In this limit $\Omega_{q\bar{q}}$
can be represented as a power series in $M$. Below we give
explicit expressions for the sigma model.
Taking into account that $\Omega_{q\bar{q}}$ is an even function of $M$,
one can write
\begin{equation}
\Omega_{q\bar{q}}(T,\mu;M)=\Omega_0(T,\mu)+\frac{M^2}{2}\left (
\frac{\partial^2 \Omega_{q\bar{q}}}{\partial M^2}\right )_{M=0}+...
\label{omqq}
\end{equation}
Here the first term, $\Omega_0(T,\mu)\equiv \Omega_{q\bar{q}}(T,\mu,0)$,
can be easily calculated for arbitrary $T$ and $\mu$. The well-known 
result is
\begin{equation}
\Omega_0(T,\mu)=-\frac{\nu_q}{2\pi^2}\left [\frac{7\pi^4}{180}T^4+
\frac{\pi^2}{6}T^2\mu^2+\frac{1}{12}\mu^4 \right ].
\end{equation}
The quark number and entropy densities for massless fermions are obtained
by differentiating $\Omega_0(T,\mu)$ with respect to  $\mu$ and $T$, 
respectively,
\begin{equation}
n=\frac{\nu_q}{6\pi^2}(\pi^2T^2\mu+\mu^3),
\end{equation}
\begin{equation}
s=\frac{\nu_q}{6\pi^2}\left (\frac{7\pi^4}{15}T^3+\pi^2T\mu^2 \right ).
\end{equation}
The second term in eq. (\ref{omqq}) differs only by a factor of $M$ from
the scalar density defined in eq. (\ref{scaldens}). A straightforward
calculation gives
\begin{equation}
\left (\frac{\partial^2\Omega}{\partial M^2} \right )_{M=0}=
\left (\frac{\rho_s}{M} \right )_{M\rightarrow 0}=\nu_q \left (\frac{T^2}{12}
+\frac{\mu^2}{4\pi^2} \right ).
\end{equation}
This can be used to estimate the pion and sigma
masses at large $T$ and/or $\mu$.
Expressing $M^2$ in terms of mean $\pi$ and $\sigma$ fields,
eq. (\ref{mmass}), and using the definition of effective masses from
eq. (\ref{qmass}), one arrives at the following asymptotic 
($M\rightarrow 0$) expression for the pion and sigma masses
\begin{equation}
M^2_{\pi}=M^2_{\sigma}=g^2\nu_q \left ( \frac{T^2}{12}+\frac{\mu^2}{4\pi^2}
\right ).
\end{equation}
It shows that deep in the chiral symmetric phase
the pion and sigma masses are degenerate and large. At high temperatures 
($T\gg\mu$), $M_\pi=M_\sigma=
gT$, where  $g\sim 3$ in our calculations.
Therefore, the contribution of pion and sigma excitations to the 
thermodynamical potential is negligible.
%It is easy to check that for $M/T\gg1$ the contribution of
%bosonic excitations to the thermodynamic potential
%is negleagable.

In case of the NJL model the above expressions are slightly
modified due to the finite cut-off $\Lambda$ in the momentum 
integration.

\newpage

\newpage

\end{document}